\documentclass[notitlepage,aps,prl,reprint,twocolumn,longbibliography,superscriptaddress]
{revtex4-1}
\usepackage{graphicx}
\usepackage{amsmath}
\usepackage{amssymb}
\usepackage{comment}
\usepackage[colorlinks, allcolors=blue]{hyperref}
\usepackage[all]{hypcap}
\usepackage[mathlines]{lineno}
\usepackage{physics}
\usepackage{wrapfig}
\usepackage{lipsum}
\usepackage{ulem}
\usepackage{siunitx}

\newcommand{\pref}[2]{\hyperref[#1]{\ref{#1}(#2)}}
\newcommand{\preff}[2]{\hyperref[#1]{\ref{#1}#2}}
\newcommand{\eqpref}[1]{\hyperref[#1]{(\ref{#1})}}

\newcommand{\squig}{{\raise.17ex\hbox{$\scriptstyle\sim$}}}

\begin{document}
	\title{Strongly interacting Rydberg atoms in synthetic dimensions with a magnetic flux}
	\author{Tao Chen}
        \thanks{These authors contributed equally to this work.}
        \author{Chenxi Huang}
        \thanks{These authors contributed equally to this work.}
	\affiliation{Department of Physics, University of Illinois at Urbana-Champaign, Urbana, IL 61801-3080, USA}
	\author{Ivan Velkovsky}
	\affiliation{Department of Physics, University of Illinois at Urbana-Champaign, Urbana, IL 61801-3080, USA}
	\author{Kaden~R.~A.~Hazzard}
        \email{kaden@rice.edu}
        \affiliation{Department of Physics and Astronomy, Rice University, Houston, TX 77005, USA}
        \affiliation{Rice Center for Quantum Materials, Rice University, Houston, TX 77005, USA}
        \affiliation{Department of Physics, University of California, Davis, CA 95616, USA}
	\author{Jacob P. Covey}
	\email{jcovey@illinois.edu}
	\affiliation{Department of Physics, University of Illinois at Urbana-Champaign, Urbana, IL 61801-3080, USA}
	\author{Bryce Gadway}
	\email{bgadway@illinois.edu}
	\affiliation{Department of Physics, University of Illinois at Urbana-Champaign, Urbana, IL 61801-3080, USA}
	\date{\today}
\begin{abstract}
Synthetic dimensions, wherein dynamics occurs in a set of internal states, have found great success in recent years in exploring topological effects in cold atoms and photonics.
However, the phenomena thus far explored have largely been restricted to the non-interacting or weakly interacting regimes. Here, we extend the synthetic dimensions playbook to strongly interacting systems of Rydberg atoms prepared in optical tweezer arrays.
We use precise control over driving microwave fields to introduce a tunable $U(1)$ flux in a four-site lattice of coupled Rydberg levels.
We find highly coherent dynamics, in good agreement with theory. Single atoms show oscillatory dynamics controllable by the gauge field. Small arrays of interacting atoms exhibit behavior suggestive of the emergence of ergodic and arrested dynamics in the regimes of intermediate and strong interactions, respectively.
These demonstrations pave the way for future explorations of strongly interacting dynamics and many-body phases in Rydberg synthetic lattices.
\end{abstract}

\maketitle

Analog quantum simulation in atomic, molecular, and optical systems has seen tremendous growth over the past decades. Recently, a flurry of activity has expanded analog simulations through synthetic dimensions~\cite{Ozawa2019,Yuan:18,Hazz-SynthDim-Rev}, where dynamics occurs not in space but in alternative degrees of freedom such as spin.
Since the first proposals a decade ago~\cite{Boada,celi2014synthetic}, the synthetic dimensions approach has permeated photonic and atomic physics experiment, with demonstrations in systems of atomic hyperfine states~\cite{Sthul2015,Mancini2015}, metastable atomic ``clock'' states~\cite{Wall-Opt,Livi-Synth,Kolkowitz2017}, atomic momentum states~\cite{Gadway-KSPACE,Meier-AtomOptics,Chen2021}, trap states~\cite{Price-Shaking,Barontini-Bloch}, photonic frequency modes~\cite{synth-freq}, orbital angular momentum modes~\cite{Cardano2017}, time-bin modes~\cite{Hafezi-TimeBin}, and more.
The realization of synthetic dimensions in these diverse platforms has led to a plethora of new simulation capabilities~\cite{Ozawa2019,Yuan:18}. However, studies have been almost entirely restricted to the non-interacting regime, with just a handful probing collective mean-field interactions in synthetic dimensions~\cite{an2018corr,Bromley2018,Xie-Walk,An-nonlinear,Wang-Interactions,Price-photons} and only one recent report of strongly correlated dynamics in synthetic dimensions~\cite{zhou2022observation}.

\begin{figure}[t!]
	\includegraphics[width=0.95\columnwidth]{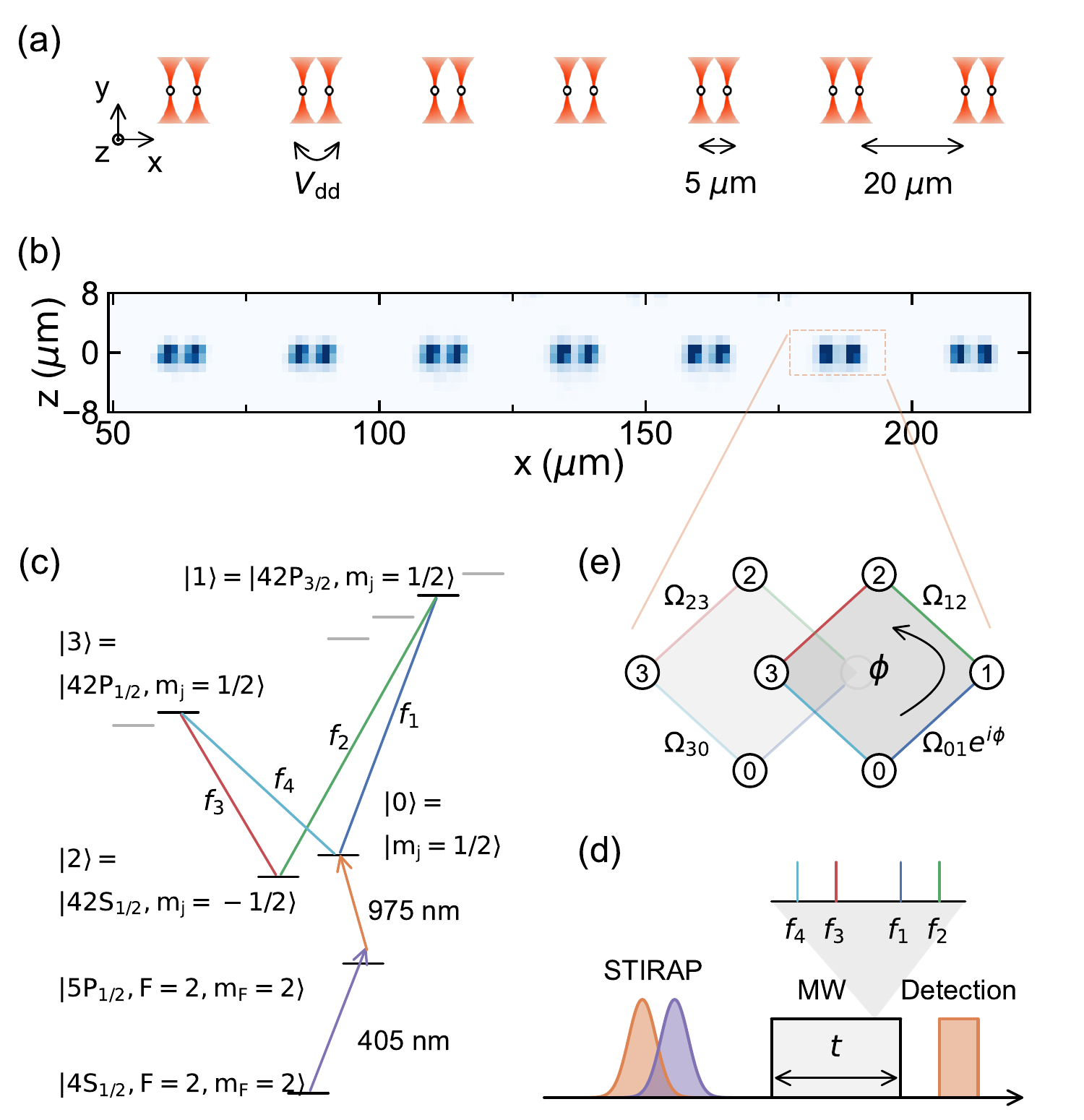}
	\caption{\textbf{Rydberg synthetic dimensions for tweezer-trapped atom arrays.}
        \textbf{(a)}~A pairwise array of optical tweezer traps is used to initialize isolated pairs of $^{39}$K atoms.
        %for the study of strongly interacting particles in a Rydberg synthetic lattice.
        \textbf{(b)}~Averaged (over $1000$~shots) fluorescence image of the atom array. 
        \textbf{(c-d)}~Ground state atoms are transferred via STIRAP to the $|42 S_\textrm{1/2}, m_J = 1/2\rangle$ Rydberg level and then exposed to microwaves (with frequencies $f_{1-4}$) to simultaneously drive multiple transitions between Rydberg levels. Dynamics of the Rydberg state populations is achieved by state-selective depumping and the imaging of ground state atoms. 
        \textbf{(e)}~The engineered synthetic Rydberg lattice, a diamond plaquette with flux $\phi$ that is tuned via the microwave phases.
        %After a duration $t$ of microwave-driven dynamics, readout is accomplished by state-selective depumping to the ground state followed by fluorescence imaging.
}
\label{FIG:fig1}
\end{figure}

Several years ago, arrays of trapped molecules and Rydberg atoms were proposed~\cite{Sundar2018,SundarPRA,Membranes} as an alternative paradigm for exploring synthetic dimensions with strong interactions. In this approach, one starts with a dipolar spin system in which interactions naturally play a significant role~\cite{Yan2013,Browaeys_2016,Gadway_2016}. Then, by introducing tailored microwaves that drive transitions between internal states in a way that mimics the hopping structure of a lattice tight-binding model, the spin system is transformed into a playground for exploring the dynamics of strongly interacting matter in a synthetic dimension.
In the past year, the team of Kanunga and co-workers have demonstrated the first Rydberg synthetic lattice~\cite{Kanungo2022}, engineering and probing topological band structures formed from the Rydberg levels of individual Sr atoms. While this demonstration~\cite{Kanungo2022} has laid the foundation for future developments of Rydberg and molecular synthetic lattices~\footnote{see also Ref.~\cite{Blackmore} for steps towards molecular synthetic dimensions, as well as related early work in Rydbergs and molecules~\cite{Signoles2014,floss2015observation}}, it lacked the key ingredient motivating the use of Rydberg atoms: strong dipole-dipole interactions.

\begin{figure}[t!]
	\includegraphics[width=1\columnwidth]{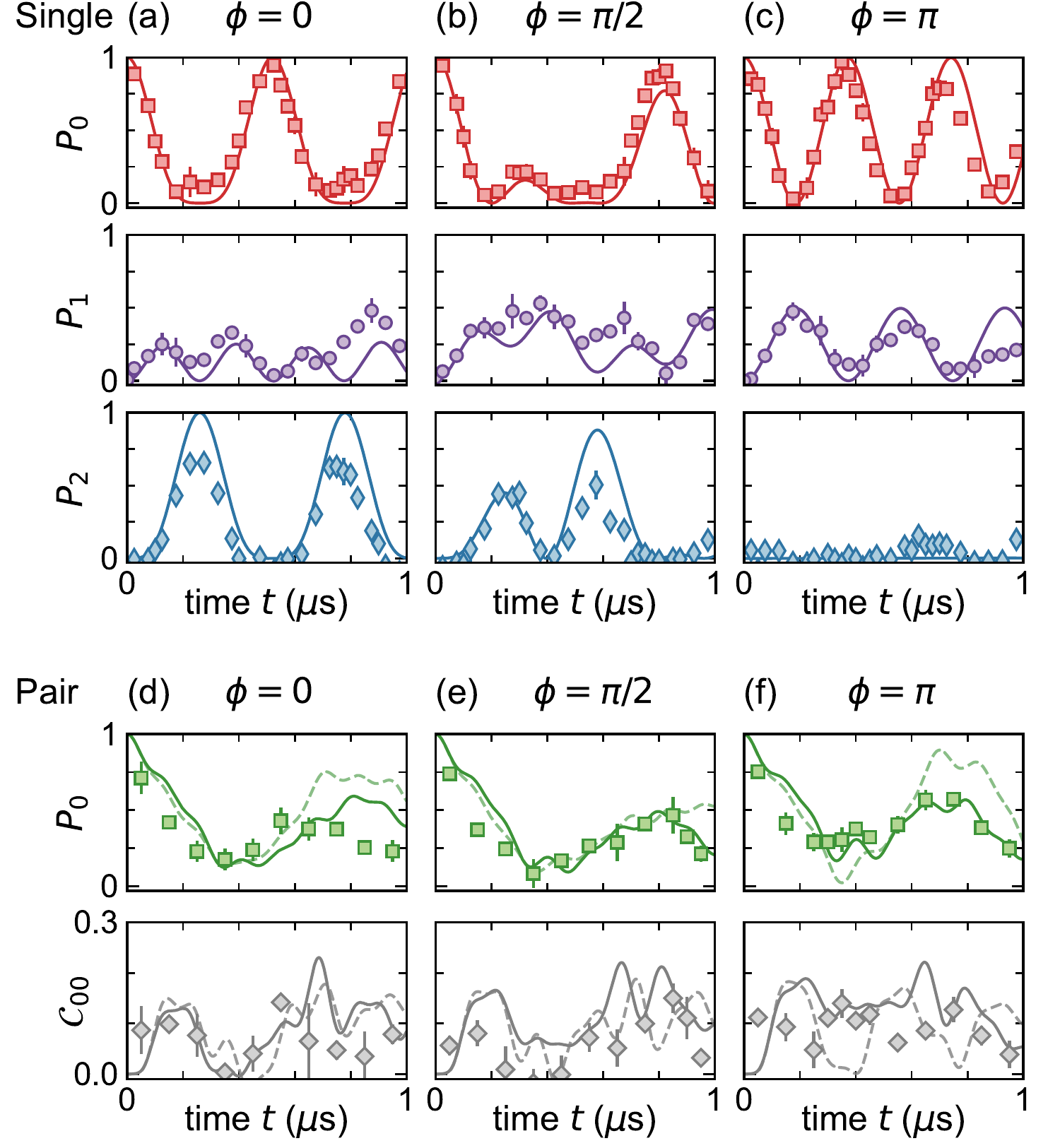}
	\caption{\textbf{Dynamics of Rydberg atoms and pairs of atoms in a synthetic flux plaquette.}
        \textbf{(a-c)}~For flux $\phi$ of (a) $0$, (b) $\pi/2$, and (c) $\pi$, we plot the average populations of single atoms at the synthetic sites $\ket{0}$, $\ket{1}$, and $\ket{2}$ (top to bottom).
        \textbf{(d-f)}~For the same flux values as above, but for the case of interacting atom pairs, we plot (top) the average atom population at site $\ket{0}$, $P_0$, and (bottom) the two-atom correlator $C_{00}$.
        The error bars for all data are the standard error of multiple independent data sets taken under the same condition. The typical singles dataset is derived from roughly 200 post-selected images while pairs relate to roughly 50 post-selected images.
        In (a-c), the theory curves are the ideal dynamics of Eq.~\ref{eq:Ham} with no free parameters ($\Omega/h = 1.92(6)$~MHz). For (d-e), the solid line is the theory for Eq.~\ref{eq:Ham} combined with all expected  dipolar interaction terms ($\Omega/h = 1.92(6)$~MHz, $V/h = 3.44(8)$~MHz). For comparison, the dashed lines neglect the state-changing dipolar contributions~\cite{SuppMats}.
}
\label{FIG:fig2}
\end{figure}

In this paper, we extend the capabilities of Rydberg synthetic dimensions by engineering an internal-state lattice with a tunable artificial gauge field~\cite{An-FluxLadder,FluxNonrecip,tHooft,Liang-Harp-Hof,Nasc-Laughlin,Yong-Flux-Cylinder} for small arrays of strongly interacting atoms~\cite{Browaeys2020-NatPhys}.
We show that the promising results of Ref.~\cite{Kanungo2022}, wherein continuous microwave coupling is performed for single Rydberg atoms excited from a bulk sample, extend directly to the real-time dynamical control of atoms prepared in optical tweezer arrays~\cite{Browaeys2020-NatPhys,Kaufman2021}. The control of the artificial gauge field in the synthetic dimension follows naturally from our phase-coherent control of the driving microwave fields.
Finally, strong nearest-neighbor interactions in the synthetic dimension lead to strong modifications of the population dynamics as well as the observation of atom-atom correlations.
%Strong interactions effects in two- and few-atom arrays, even the observation of atom-atom correlations.
This work paves the way for future explorations of strongly-correlated dynamics and phases of matter in Rydberg and molecular synthetic dimensions.

Our experiments begin by probabilistically loading $^{39}$K atoms~\cite{Gross-SciPost,JacksonPRR} into optical tweezer arrays as depicted in Fig.~\pref{FIG:fig1}{a,b}, nondestructively imaging the atoms for subsequent post-selection, and cooling the atoms by gray molasses~\cite{JacksonPRR,Salomon_2013}. We optically pump the atoms (with a quantization $B$-field of $\sim$27~G along the $z$ axis) to a single ground level $\ket{4 S_\textrm{1/2}, F=2, m_F = 2}$ with an efficiency of $\sim$98(1)$\%$, and then we further cool the atoms by trap decompression to $\sim$4~$\mu$K. We then suddenly turn off the confining tweezer trap.

The atoms undergo a fixed free release time of 5~$\mu$s, during which all of the dynamics in the Rydberg synthetic lattice occurs.
The atoms are promoted to an initial Rydberg level, undergo microwave-driven dynamics between Rydberg levels, and are de-excited in a manner that allows for Rydberg state-specific readout. Following de-excitation, ground state atoms are recaptured in the trap and imaged with high fidelity. Atoms remaining in the Rydberg levels are weakly anti-trapped by the tweezers, and are effectively lost between the initial and final images. This bright/dark discrimination between ground and Rydberg levels, combined with state-selective de-excitation, allows us to study the state-resolved dynamics of the Rydberg level populations.

The initial excitation to the Rydberg level $\ket{0}\equiv \ket{42 S_\textrm{1/2}, m_J = 1/2}$ is accomplished via two-photon (``lower leg,'' $\sim$405~nm, and ``upper leg,'' $\sim$975~nm)
%\textcolor{green}{JPC: define STIRAP? Reference Browaeys?}
stimulated Raman adiabatic passage (STIRAP) via the $\ket{5 P_\textrm{1/2}, F = 2, m_F = 1/2}$ intermediate state~\cite{Cubel-STIRAP,sylvain18}.
The averaged one-way STIRAP efficiency is $\sim$94(1)\%~\cite{SuppMats}.
After populating this initial state, we turn on a set of microwave tones that allow atoms to ``hop'' between the sites of an effective lattice in the ``synthetic dimension'' spanned by the Rydberg levels~\cite{Boada,Sundar2018,Kanungo2022}.

As shown in Fig.~\pref{FIG:fig1}{c-e}, we identify the sites of the synthetic Rydberg lattice with the atomic Rydberg levels as $\ket{0}\equiv \ket{42 S_\textrm{1/2}, m_J = 1/2}$, $\ket{1}\equiv \ket{42 P_\textrm{3/2}, m_J = 1/2}$, $\ket{2}\equiv \ket{42 S_\textrm{1/2}, m_J = -1/2}$, and $\ket{3}\equiv \ket{42 P_\textrm{1/2}, m_J = 1/2}$.
A single flux plaquette is formed by adding microwave tones that resonantly drive four pairwise transitions within this set of states.

The effective single-atom Hamiltonian is given by
\begin{align}
H = \sum_{\langle i,j\rangle}
\frac{\Omega_{ij}}{2} \ \hat{c}_{j}^\dagger \hat{c}_{i} + {\rm h.c.} = \frac{\Omega}{2} \sum_{\langle i,j\rangle}
e^{i\phi_{ij}} \ \hat{c}_{j}^\dagger \hat{c}_{i} + {
\rm h.c.} \ , \label{eq:Ham}
\end{align}
where the nearest-neighbor tunneling terms are related to the amplitudes ($A_i$) and phases ($\varphi_{i}$, at the atoms) of the different microwave tones $f_i$ as $\Omega_{01} \propto A_1 e^{i\varphi_1}$, $\Omega_{12} = \Omega^*_{21} \propto A_2 e^{-i\varphi_2}$, $\Omega_{23} \propto A_3 e^{i\varphi_3}$, and $\Omega_{30} = \Omega^*_{03} \propto A_4 e^{-i\varphi_4}$. 
The magnitudes of these nearest-neighbor hopping terms are calibrated based on pairwise Rabi dynamics~\cite{SuppMats} and are set to a common value $\Omega$.
The relative phase of each tone at the atoms is controllable by the source phase, and in particular we set the overall plaquette flux $\phi$ via the source phase of the $f_1$ tone.

Figure~\ref{FIG:fig2} displays the dynamics of the state populations (starting from the $\ket{0}$ state at $t=0$). The populations are corrected for measurement errors related to the STIRAP infidelity and Rydberg-vs.-ground discrimination infidelity~\cite{SuppMats}.
%The primary readout of the atomic Rydberg levels is performed by fast depumping of the $\ket{0}$ state by application of the ``upper leg'' STIRAP laser.
The $\ket{0}$ state is measured by direct depumping by the ``upper leg'' STIRAP laser after some evolution time. To access the $\ket{1}$ state, which shares identical population dynamics in this model as the $\ket{3}$ state, we first apply a $\pi$ pulse on the $\ket{0}$  to $\ket{1}$ transition prior to depumping. To access the $\ket{2}$ state, which is quite close in energy to the $\ket{0}$ state, we simply apply a strong (high-bandwidth) depumping pulse to measure the combined population of $\ket{0}$ and $\ket{2}$, $P_{0+2}$. We then extract the $\ket{2}$ state population as $P_2 = P_{0+2} - P_0$.
For single atoms we generally find good agreement with the population dynamics for the examined flux values of Fig.~\ref{FIG:fig2}(a)~$0$, (b)~$\pi/2$, and (c)~$\pi$. The changing timescales for $P_0$ recurrence reflects the flux-tuned spectral gaps of the plaquette energy spectrum. One stark signature seen in Fig.~\pref{FIG:fig2}{c}, for $\pi$ flux, is the absence of population appearing at state $\ket{2}$, which results from destructive interference of the clockwise and counterclockwise pathways.

\begin{figure}[t!]
	\includegraphics[width=\columnwidth]{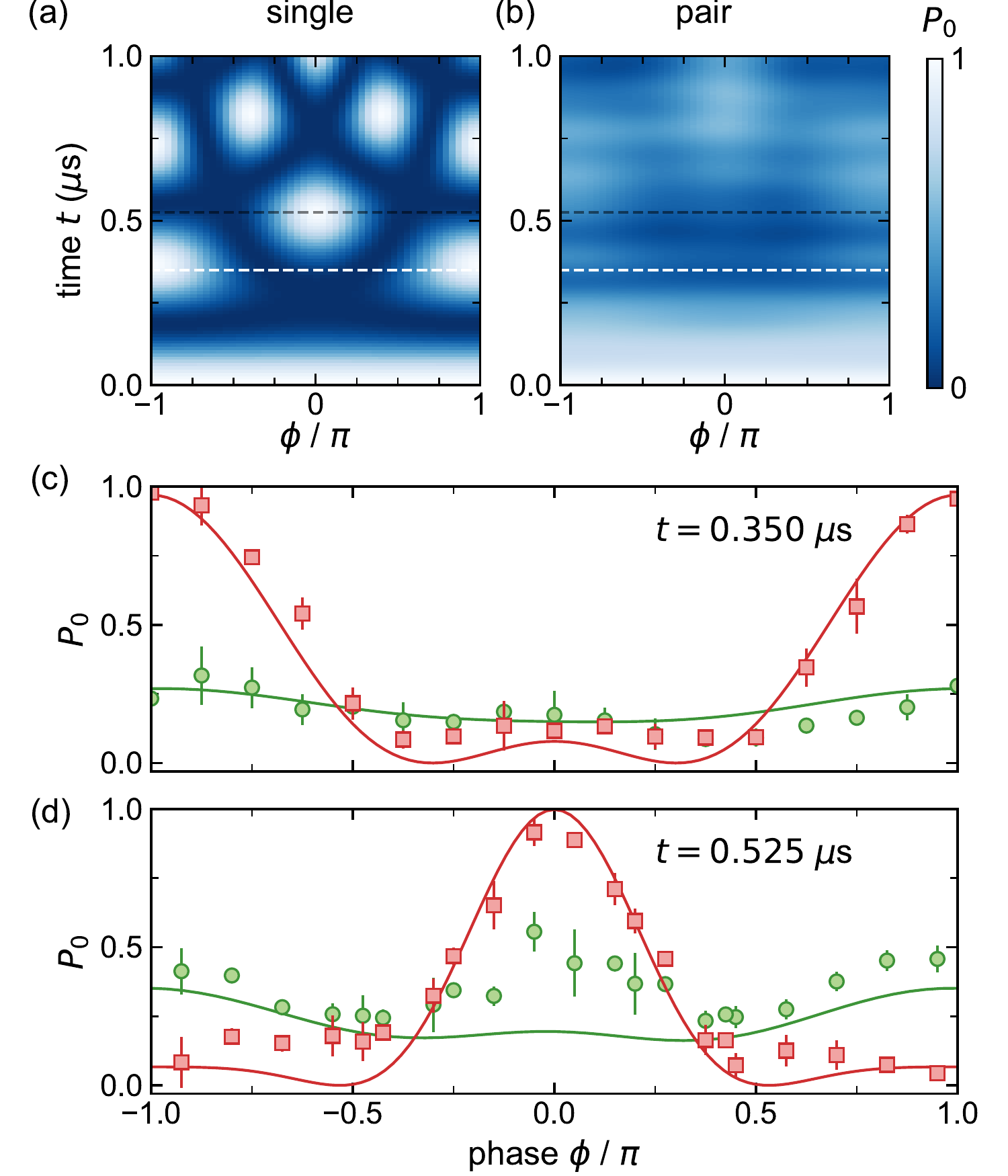}
	\caption{\textbf{Flux-dependence of atom and atom pair dynamics.}
        \textbf{(a)}~A plot of the average probability vs. time $t$ and flux $\phi$ for atoms initialized at state $\ket{0}$ to remain at that state ($P_0$, for single atoms). 
        \textbf{(b)}~The same quantity as plotted in (a), and with a common color bar at right, but as calculated for pairs of atoms interacting via dipole-dipole interactions as described in the text.
        \textbf{(c,d)}~Measured $P_0$ for single atoms (red squares) and pairs (green circles) after evolution times of $t = 0.350$~$\mu$s and $0.525$~$\mu$s, shown along with theory curves (solid lines), corresponding to cuts along the white and black dashed lines in panels (a,b).
        The error bars in (c,d) are the standard error of multiple independent data sets.
}
\label{FIG:fig3}
\end{figure}

The dynamics of lone atoms in Fig.~\pref{FIG:fig2}{a-c} verifies our faithful implementation of the single-particle synthetic lattice and flux control. In Fig.~\pref{FIG:fig2}{d,e}, we use isolated pairs of atoms to investigate how strong inter-particle interactions enrich the dynamics. The principal interactions between Rydberg atoms in this system involve long-ranged ($1/r^3$, with $r$ the inter-particle spacing) dipolar exchange~\cite{Browaeys_2016}.
In our system, having a uniform quantization axis aligned at an angle $\theta = \pi/2$ with respect to the displacement vectors between pairs of atoms, the primary interactions to consider are 
resonant dipolar-exchange terms of the form $\ket{i}_A \ket{j}_B \leftrightarrow \ket{i}_B \ket{j}_A$, or ``flip-flop'' interactions, in which the synthetic location of internal Rydberg states $|i\rangle$ and $|j\rangle$ are swapped between the atoms at positions labeled $A$ and $B$ (for nearest neighbors $\langle i,j \rangle$, with  $i,j \in \{ 0,1,2,3\}$), but the net populations of the Rydberg levels are conserved. These $\Delta \ell=0$ dipolar terms that conserve the net internal angular momentum (and its projection along the quantization axis) also naturally conserve the total energy in a spatially uniform system, and thus result in resonant exchange dynamics~\cite{Yan2013,Sylvain}.
In our system, for pairs of atoms spaced at a distance of 5~$\mu$m [Fig.~\pref{FIG:fig1}{a,b}], the resonant dipolar exchange energies can be enumerated as $\{V_{01},V_{12},V_{23},V_{30}\} \approx \{2,-0.5,1,-1\}V$, where $V/h = 3.44(8)$~MHz~\cite{SuppMats}.
Because we operate at a modest magnetic field and with relatively strong interactions, additional off-resonant state-changing dipolar interaction terms ($\Delta \ell=\pm 2$, not conserving the net internal angular momentum or the individual state populations) also influence the state population dynamics~\cite{SuppMats}.

For pairs, we restrict ourselves to measuring the population of $\ket{0}$ for each atom, as the basis rotation pulses used for the readout of other internal states are influenced by the presence of strong interactions.
Figure~\pref{FIG:fig2}{d} shows the average probability for a pair of atoms to reside at the site $\ket{0}$. We compare to no-free-parameter simulations of Eq.~\ref{eq:Ham}, also incorporating the full set of expected interactions (solid line). For comparison, we also show simulations (dashed lines) that ignore the state-changing dipolar terms, which can be suppressed by operating at larger magnetic bias fields or with larger inter-atomic spacings.
For pairs in this intermediate interaction regime [$V/\Omega = 1.8(1)$], we observe that the dipolar interactions strongly modify the dynamics, in general increasing the dynamical timescales and decreasing the amplitude of recurrences. As a more direct probe of interaction-driven correlations, we measure the two-atom correlator
%$C_{00} = \langle P_{0,L} P_{0,R} \rangle - \langle P_{0,L}\rangle \langle P_{0,R}\rangle$, 
$\mathcal{C}_{00} = \langle \hat{c}^\dagger_{0, L}\hat{c}_{0,L}\hat{c}^\dagger_{0,R}\hat{c}_{0,R}\rangle - \langle \hat{c}^\dagger_{0,L}\hat{c}_{0,L}\rangle \langle\hat{c}^\dagger_{0,R}\hat{c}_{0,R}\rangle$
with $L$ and $R$ referring to the left and right atoms of isolated pairs. This quantity vanishes in the absence of interactions, and grows as the atoms develop correlations of their positions in the synthetic lattice. Both the $P_{0}$ and $C_{00}$ dynamics are in good agreement with our textbook theory expectations, confirming that dipolar Rydberg atom arrays are a promising platform for exploring coherent interactions in tunable synthetic lattices.
%Additionally, our experimental setup admits state-changing dipolar exchange terms that do not conserve the net internal angular momentum ($q=\pm 2$), and in fact lead to atoms leaving the synthetic lattice ($\{ 0,1,2,3\}$) subspace. For our moderate magnetic field strengths, these terms are weakly off-resonant and thus suppressed, but still play some role in these experiments 
%To note, for the case of doubles, the readout of different Rydberg levels through fast $\pi$ pulses prior to depumping is challenged by the large influence of interactions (and technical limitations on the Rabi rates presently achievable on the various transitions). Hence, for doubles and beyond we herein restrict ourselves to measuring the population in the $\ket{0}$ state.

We now more thoroughly explore in Fig.~\ref{FIG:fig3} the flux-dependent dynamics for individual atoms and atom pairs. Figure~\pref{FIG:fig3}{a,b} show numerical simulations of the full flux-dependence of the $P_0$
%\textcolor{green}{JPC: This `bar' notation does not seem to be showing up properly.}
dynamics for singles and pairs. For singles, as described before, the changing timescales for recurrences of the measure $P_0$ simply reflect the flux-modified gaps of the system's energy spectrum. For our measurements, we probe precisely at the first expected $P_0$ recurrence time for singles at flux values of $\phi = \pi$ and $0$, namely at $t = 0.350$~$\mu$s in Fig.~\pref{FIG:fig3}{c} and $t = 0.525$~$\mu$s in Fig.~\pref{FIG:fig3}{d}, respectively. For singles, we observe good general agreement with the full flux-dependence of the expected $P_0$ dynamics.
For doubles, we observe both in theory and experiment that the dynamics slow down considerably, such that $P_{0}$ remains relatively small at the single-atom recurrence times.
Interestingly, one finds already for this intermediate interaction regime [$V/\Omega = 1.8(1)$] that the pair dynamics for a flux of 0 and $\pi$ look somewhat similar, suggestive of the expected response in the strong interaction limit where mobile bound pairs~\cite{Preiss-QWalk} would display an enhanced flux sensitivity.
%Indeed, in the strong interaction limit, where one would expect atoms to move as a bound pair~\cite{Preiss-QWalk}, the pair dynamics should become $\pi$-periodic due to an enhanced flux sensitivity.
%Even for these fairly moderate interactions, one finds that the dynamics of doubles is effectivley $\pi$-periodic with the flux $\phi$.

\begin{figure}[t!]
	\includegraphics[width=\columnwidth]{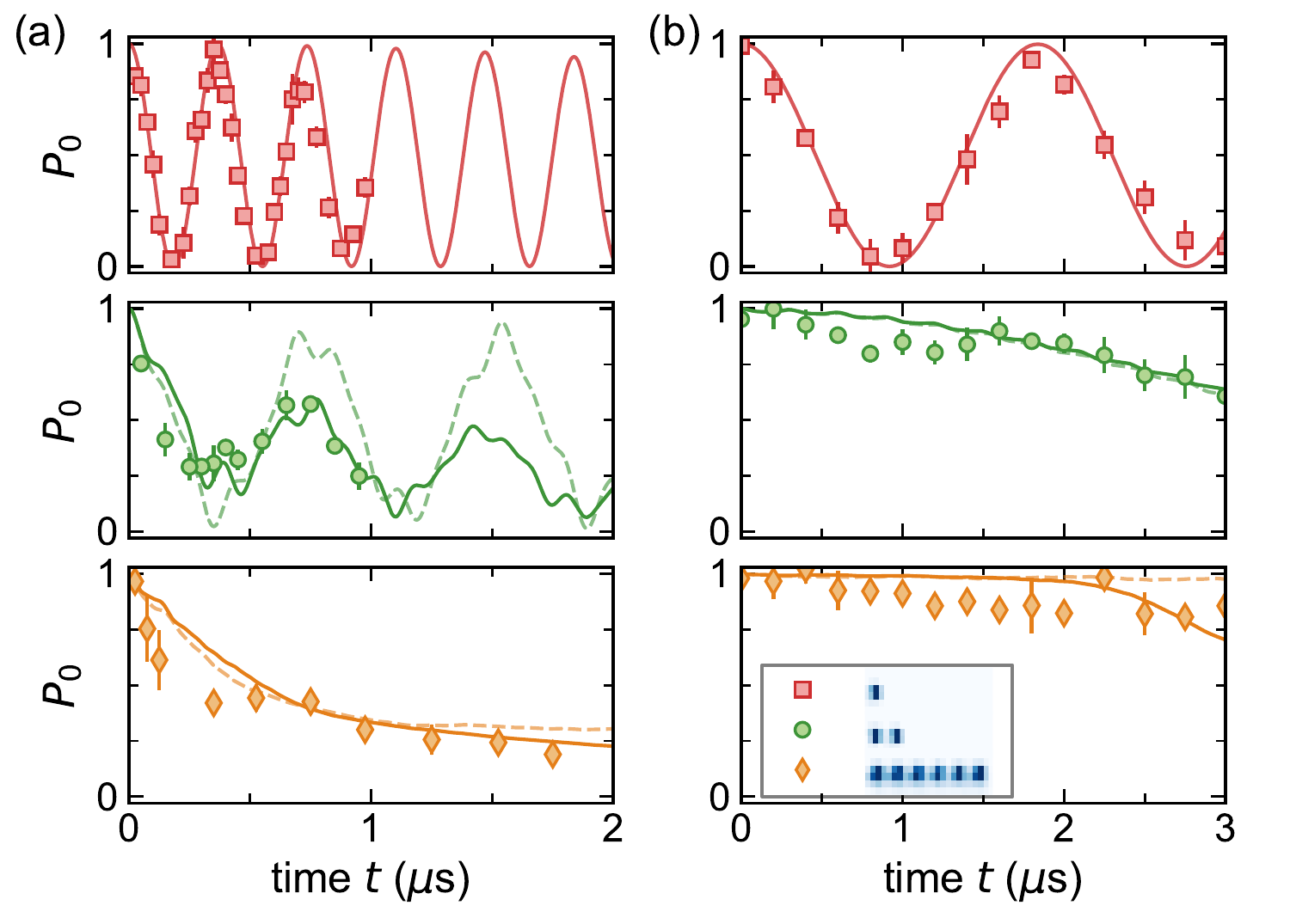}
        \centering
	\caption{\label{FIG:fig4}
        \textbf{Scrambling and self-trapping in few-atom arrays.}
        \textbf{(a)}~Dynamics for single atoms (red squares), atom pairs (green circles), and six-atom arrays (orange diamonds) under a flux of $\pi$ and for a characteristic interaction-to-hopping ratio of $V/\Omega = 1.8(1)$ with $\Omega/h = 1.92(6)$~MHz. The singles and pairs data are the same as for Fig.~\pref{FIG:fig2}{c,f}.
        \textbf{(b)}~Same quantities as in (a), but for reduced hopping amplitude $\Omega/h = 0.38(1)$ MHz and $V/\Omega = 9.0(5)$.
        The solid (dashed) lines in panels (a,b) are the no-free-parameters theory based on Eq.~\ref{eq:Ham} and the full set (resonant-only set) of expected dipolar interactions.
        Experimental error bars are the standard error from multiple independent data sets.
}
\end{figure}

Finally, we explore how interactions in Rydberg synthetic dimensions can have an even richer influence on the dynamics as we extend towards many-atom arrays. In Fig.~\pref{FIG:fig4}{a,b}, we contrast the $\phi = \pi$ dynamics of one, two, and six-atom arrays for intermediate [(a), $V/\Omega = 1.8(1)$, $\Omega/h = 1.92(6)$~MHz] and large [(b), $V/\Omega = 9.0(5)$, $\Omega/h = 0.38(1)$~MHz] interaction-to-tunneling ratios. For both cases, $P_0$ oscillates with high coherence and a single frequency for single atoms (there is only a single energy gap value at $\pi$ flux).
However, interactions lead to qualitatively different dynamics in multi-atom arrays~\cite{SuppMats}.
In Fig.~\pref{FIG:fig4}{a}, for $V/\Omega \sim 1.8$, the macroscopic observable $P_0$ shows coherent revivals with a structured time-dependence for pairs, and less oscillations but a clear decay for six-atom clusters. Specifically, numerical simulations for the six-atom array show a rapid relaxation to $P_0 \approx 1/4$, suggestive of an approach to ergodicity in this closed many-body system. At very long times, the deviation of the numerics from $1/4$ result simply from state non-conserving interactions~\cite{SuppMats}.
The dynamics of arrays relative to singles changes remarkably for strong interactions, $V/\Omega \sim 9$, as shown in Fig.~\pref{FIG:fig4}{b}.
For pairs, we observe only a very slow decay of $P_0$ over the 3~$\mu$s measurement window, consistent with the prediction of pair-hopping in this large $V/\Omega$ limit~\cite{Preiss-QWalk}. The $P_0$ dynamics is even slower for six-atom clusters, and in this case the dynamics should be attributed almost entirely to state non-conserving dipolar processes. In the case of only resonant interactions (dashed line), a full interaction-driven immobilization or self-trapping is expected in this strong interaction regime, related to the emergence of quantum strings~\cite{Sundar2018,SundarPRA,Membranes}. In this large $V/\Omega$ regime, one would expect the system to be prone to Hilbert space fragmentation~\cite{Frag1,Frag2} and fully arrested dynamics under added perturbations (\textit{e.g.}, a gradient or disorder).

%While only the onset of self-trapping is observable in these measurements, future experiments promise to explore the predicted formation and stability of quantum strings and membranes in dipolar atom arrays
These observations of highly-excited self-trapped strings pave the way for future experiments to explore the predicted ground state quantum string and membrane phases in Rydberg synthetic dimensions~\cite{Sundar2018,SundarPRA,Membranes}. Beyond the physics of dipolar strings, which arise naturally for the most generic synthetic lattices, the excellent coherence properties observed in this first exploration of tweezer-array based Rydberg synthetic dimensions bodes well for extensions to study the interplay of interactions and topology or frustration in more complex synthetic lattices, such as flat-band models, extended flux lattices, and models with tunable disorder and dissipation.

\section{Acknowledgements}
This material is based upon work supported by the National Science Foundation under grant No.~1945031 and the AFOSR MURI program under agreement number FA9550-22-1-0339.  K. R. A. H. acknowledges support from the Robert A. Welch Foundation (C-1872), the National Science Foundation (PHY-1848304), the Office of Naval Research (N00014-20-1-2695), and the W. F. Keck
 Foundation (Grant No. 995764).  K. R. A. H.'s contribution benefited from discussions at the Aspen Center for Physics, supported by the National Science Foundation grant PHY-1066293, and the KITP, which was supported in part by the National Science Foundation under Grant No. NSF PHY-1748958.
We thank A.~X.~El-Khadra and P.~Draper for early stimulating discussions. We also acknowledge Jackson Ang'ong'a for early work contributing to the experimental apparatus.

\bibliographystyle{apsrev4-1}
\bibliography{Ryd}

\begin{thebibliography}{57}%
\makeatletter
\providecommand \@ifxundefined [1]{%
 \@ifx{#1\undefined}
}%
\providecommand \@ifnum [1]{%
 \ifnum #1\expandafter \@firstoftwo
 \else \expandafter \@secondoftwo
 \fi
}%
\providecommand \@ifx [1]{%
 \ifx #1\expandafter \@firstoftwo
 \else \expandafter \@secondoftwo
 \fi
}%
\providecommand \natexlab [1]{#1}%
\providecommand \enquote  [1]{``#1''}%
\providecommand \bibnamefont  [1]{#1}%
\providecommand \bibfnamefont [1]{#1}%
\providecommand \citenamefont [1]{#1}%
\providecommand \href@noop [0]{\@secondoftwo}%
\providecommand \href [0]{\begingroup \@sanitize@url \@href}%
\providecommand \@href[1]{\@@startlink{#1}\@@href}%
\providecommand \@@href[1]{\endgroup#1\@@endlink}%
\providecommand \@sanitize@url [0]{\catcode `\\12\catcode `\$12\catcode
  `\&12\catcode `\#12\catcode `\^12\catcode `\_12\catcode `\%12\relax}%
\providecommand \@@startlink[1]{}%
\providecommand \@@endlink[0]{}%
\providecommand \url  [0]{\begingroup\@sanitize@url \@url }%
\providecommand \@url [1]{\endgroup\@href {#1}{\urlprefix }}%
\providecommand \urlprefix  [0]{URL }%
\providecommand \Eprint [0]{\href }%
\providecommand \doibase [0]{http://dx.doi.org/}%
\providecommand \selectlanguage [0]{\@gobble}%
\providecommand \bibinfo  [0]{\@secondoftwo}%
\providecommand \bibfield  [0]{\@secondoftwo}%
\providecommand \translation [1]{[#1]}%
\providecommand \BibitemOpen [0]{}%
\providecommand \bibitemStop [0]{}%
\providecommand \bibitemNoStop [0]{.\EOS\space}%
\providecommand \EOS [0]{\spacefactor3000\relax}%
\providecommand \BibitemShut  [1]{\csname bibitem#1\endcsname}%
\let\auto@bib@innerbib\@empty
%</preamble>
\bibitem [{\citenamefont {Ozawa}\ and\ \citenamefont
  {Price}(2019)}]{Ozawa2019}%
  \BibitemOpen
  \bibfield  {author} {\bibinfo {author} {\bibfnamefont {T.}~\bibnamefont
  {Ozawa}}\ and\ \bibinfo {author} {\bibfnamefont {H.~M.}\ \bibnamefont
  {Price}},\ }\href {\doibase 10.1038/s42254-019-0045-3} {\bibfield  {journal}
  {\bibinfo  {journal} {Nature Reviews Physics}\ }\textbf {\bibinfo {volume}
  {1}},\ \bibinfo {pages} {349} (\bibinfo {year} {2019})}\BibitemShut {NoStop}%
\bibitem [{\citenamefont {Yuan}\ \emph {et~al.}(2018)\citenamefont {Yuan},
  \citenamefont {Lin}, \citenamefont {Xiao},\ and\ \citenamefont
  {Fan}}]{Yuan:18}%
  \BibitemOpen
  \bibfield  {author} {\bibinfo {author} {\bibfnamefont {L.}~\bibnamefont
  {Yuan}}, \bibinfo {author} {\bibfnamefont {Q.}~\bibnamefont {Lin}}, \bibinfo
  {author} {\bibfnamefont {M.}~\bibnamefont {Xiao}}, \ and\ \bibinfo {author}
  {\bibfnamefont {S.}~\bibnamefont {Fan}},\ }\href {\doibase
  10.1364/OPTICA.5.001396} {\bibfield  {journal} {\bibinfo  {journal} {Optica}\
  }\textbf {\bibinfo {volume} {5}},\ \bibinfo {pages} {1396} (\bibinfo {year}
  {2018})}\BibitemShut {NoStop}%
\bibitem [{\citenamefont {Hazzard}\ and\ \citenamefont
  {Gadway}(2023)}]{Hazz-SynthDim-Rev}%
  \BibitemOpen
  \bibfield  {author} {\bibinfo {author} {\bibfnamefont {K.~R.~A.}\
  \bibnamefont {Hazzard}}\ and\ \bibinfo {author} {\bibfnamefont
  {B.}~\bibnamefont {Gadway}},\ }\href {\doibase 10.1063/PT.3.5225} {\bibfield
  {journal} {\bibinfo  {journal} {Physics Today}\ }\textbf {\bibinfo {volume}
  {76}},\ \bibinfo {pages} {62} (\bibinfo {year} {2023})}\BibitemShut {NoStop}%
\bibitem [{\citenamefont {Boada}\ \emph {et~al.}(2012)\citenamefont {Boada},
  \citenamefont {Celi}, \citenamefont {Latorre},\ and\ \citenamefont
  {Lewenstein}}]{Boada}%
  \BibitemOpen
  \bibfield  {author} {\bibinfo {author} {\bibfnamefont {O.}~\bibnamefont
  {Boada}}, \bibinfo {author} {\bibfnamefont {A.}~\bibnamefont {Celi}},
  \bibinfo {author} {\bibfnamefont {J.~I.}\ \bibnamefont {Latorre}}, \ and\
  \bibinfo {author} {\bibfnamefont {M.}~\bibnamefont {Lewenstein}},\ }\href
  {\doibase 10.1103/PhysRevLett.108.133001} {\bibfield  {journal} {\bibinfo
  {journal} {Phys. Rev. Lett.}\ }\textbf {\bibinfo {volume} {108}},\ \bibinfo
  {pages} {133001} (\bibinfo {year} {2012})}\BibitemShut {NoStop}%
\bibitem [{\citenamefont {Celi}\ \emph {et~al.}(2014)\citenamefont {Celi},
  \citenamefont {Massignan}, \citenamefont {Ruseckas}, \citenamefont {Goldman},
  \citenamefont {Spielman}, \citenamefont {Juzeli\ifmmode~\bar{u}\else
  \={u}\fi{}nas},\ and\ \citenamefont {Lewenstein}}]{celi2014synthetic}%
  \BibitemOpen
  \bibfield  {author} {\bibinfo {author} {\bibfnamefont {A.}~\bibnamefont
  {Celi}}, \bibinfo {author} {\bibfnamefont {P.}~\bibnamefont {Massignan}},
  \bibinfo {author} {\bibfnamefont {J.}~\bibnamefont {Ruseckas}}, \bibinfo
  {author} {\bibfnamefont {N.}~\bibnamefont {Goldman}}, \bibinfo {author}
  {\bibfnamefont {I.~B.}\ \bibnamefont {Spielman}}, \bibinfo {author}
  {\bibfnamefont {G.}~\bibnamefont {Juzeli\ifmmode~\bar{u}\else
  \={u}\fi{}nas}}, \ and\ \bibinfo {author} {\bibfnamefont {M.}~\bibnamefont
  {Lewenstein}},\ }\href {\doibase 10.1103/PhysRevLett.112.043001} {\bibfield
  {journal} {\bibinfo  {journal} {Phys. Rev. Lett.}\ }\textbf {\bibinfo
  {volume} {112}},\ \bibinfo {pages} {043001} (\bibinfo {year}
  {2014})}\BibitemShut {NoStop}%
\bibitem [{\citenamefont {Stuhl}\ \emph {et~al.}(2015)\citenamefont {Stuhl},
  \citenamefont {Lu}, \citenamefont {Aycock}, \citenamefont {Genkina},\ and\
  \citenamefont {Spielman}}]{Sthul2015}%
  \BibitemOpen
  \bibfield  {author} {\bibinfo {author} {\bibfnamefont {B.~K.}\ \bibnamefont
  {Stuhl}}, \bibinfo {author} {\bibfnamefont {H.-I.}\ \bibnamefont {Lu}},
  \bibinfo {author} {\bibfnamefont {L.~M.}\ \bibnamefont {Aycock}}, \bibinfo
  {author} {\bibfnamefont {D.}~\bibnamefont {Genkina}}, \ and\ \bibinfo
  {author} {\bibfnamefont {I.~B.}\ \bibnamefont {Spielman}},\ }\href {\doibase
  10.1126/science.aaa8515} {\bibfield  {journal} {\bibinfo  {journal}
  {Science}\ }\textbf {\bibinfo {volume} {349}},\ \bibinfo {pages} {1514}
  (\bibinfo {year} {2015})}\BibitemShut {NoStop}%
\bibitem [{\citenamefont {Mancini}\ \emph {et~al.}(2015)\citenamefont
  {Mancini}, \citenamefont {Pagano}, \citenamefont {Cappellini}, \citenamefont
  {Livi}, \citenamefont {Rider}, \citenamefont {Catani}, \citenamefont {Sias},
  \citenamefont {Zoller}, \citenamefont {Inguscio}, \citenamefont {Dalmonte},\
  and\ \citenamefont {Fallani}}]{Mancini2015}%
  \BibitemOpen
  \bibfield  {author} {\bibinfo {author} {\bibfnamefont {M.}~\bibnamefont
  {Mancini}}, \bibinfo {author} {\bibfnamefont {G.}~\bibnamefont {Pagano}},
  \bibinfo {author} {\bibfnamefont {G.}~\bibnamefont {Cappellini}}, \bibinfo
  {author} {\bibfnamefont {L.}~\bibnamefont {Livi}}, \bibinfo {author}
  {\bibfnamefont {M.}~\bibnamefont {Rider}}, \bibinfo {author} {\bibfnamefont
  {J.}~\bibnamefont {Catani}}, \bibinfo {author} {\bibfnamefont
  {C.}~\bibnamefont {Sias}}, \bibinfo {author} {\bibfnamefont {P.}~\bibnamefont
  {Zoller}}, \bibinfo {author} {\bibfnamefont {M.}~\bibnamefont {Inguscio}},
  \bibinfo {author} {\bibfnamefont {M.}~\bibnamefont {Dalmonte}}, \ and\
  \bibinfo {author} {\bibfnamefont {L.}~\bibnamefont {Fallani}},\ }\href
  {\doibase 10.1126/science.aaa8736} {\bibfield  {journal} {\bibinfo  {journal}
  {Science}\ }\textbf {\bibinfo {volume} {349}},\ \bibinfo {pages} {1510}
  (\bibinfo {year} {2015})}\BibitemShut {NoStop}%
\bibitem [{\citenamefont {Wall}\ \emph {et~al.}(2016)\citenamefont {Wall},
  \citenamefont {Koller}, \citenamefont {Li}, \citenamefont {Zhang},
  \citenamefont {Cooper}, \citenamefont {Ye},\ and\ \citenamefont
  {Rey}}]{Wall-Opt}%
  \BibitemOpen
  \bibfield  {author} {\bibinfo {author} {\bibfnamefont {M.~L.}\ \bibnamefont
  {Wall}}, \bibinfo {author} {\bibfnamefont {A.~P.}\ \bibnamefont {Koller}},
  \bibinfo {author} {\bibfnamefont {S.}~\bibnamefont {Li}}, \bibinfo {author}
  {\bibfnamefont {X.}~\bibnamefont {Zhang}}, \bibinfo {author} {\bibfnamefont
  {N.~R.}\ \bibnamefont {Cooper}}, \bibinfo {author} {\bibfnamefont
  {J.}~\bibnamefont {Ye}}, \ and\ \bibinfo {author} {\bibfnamefont {A.~M.}\
  \bibnamefont {Rey}},\ }\href {\doibase 10.1103/PhysRevLett.116.035301}
  {\bibfield  {journal} {\bibinfo  {journal} {Phys. Rev. Lett.}\ }\textbf
  {\bibinfo {volume} {116}},\ \bibinfo {pages} {035301} (\bibinfo {year}
  {2016})}\BibitemShut {NoStop}%
\bibitem [{\citenamefont {Livi}\ \emph {et~al.}(2016)\citenamefont {Livi},
  \citenamefont {Cappellini}, \citenamefont {Diem}, \citenamefont {Franchi},
  \citenamefont {Clivati}, \citenamefont {Frittelli}, \citenamefont {Levi},
  \citenamefont {Calonico}, \citenamefont {Catani}, \citenamefont {Inguscio},\
  and\ \citenamefont {Fallani}}]{Livi-Synth}%
  \BibitemOpen
  \bibfield  {author} {\bibinfo {author} {\bibfnamefont {L.~F.}\ \bibnamefont
  {Livi}}, \bibinfo {author} {\bibfnamefont {G.}~\bibnamefont {Cappellini}},
  \bibinfo {author} {\bibfnamefont {M.}~\bibnamefont {Diem}}, \bibinfo {author}
  {\bibfnamefont {L.}~\bibnamefont {Franchi}}, \bibinfo {author} {\bibfnamefont
  {C.}~\bibnamefont {Clivati}}, \bibinfo {author} {\bibfnamefont
  {M.}~\bibnamefont {Frittelli}}, \bibinfo {author} {\bibfnamefont
  {F.}~\bibnamefont {Levi}}, \bibinfo {author} {\bibfnamefont {D.}~\bibnamefont
  {Calonico}}, \bibinfo {author} {\bibfnamefont {J.}~\bibnamefont {Catani}},
  \bibinfo {author} {\bibfnamefont {M.}~\bibnamefont {Inguscio}}, \ and\
  \bibinfo {author} {\bibfnamefont {L.}~\bibnamefont {Fallani}},\ }\href
  {\doibase 10.1103/PhysRevLett.117.220401} {\bibfield  {journal} {\bibinfo
  {journal} {Phys. Rev. Lett.}\ }\textbf {\bibinfo {volume} {117}},\ \bibinfo
  {pages} {220401} (\bibinfo {year} {2016})}\BibitemShut {NoStop}%
\bibitem [{\citenamefont {Kolkowitz}\ \emph {et~al.}(2017)\citenamefont
  {Kolkowitz}, \citenamefont {Bromley}, \citenamefont {Bothwell}, \citenamefont
  {Wall}, \citenamefont {Marti}, \citenamefont {Koller}, \citenamefont {Zhang},
  \citenamefont {Rey},\ and\ \citenamefont {Ye}}]{Kolkowitz2017}%
  \BibitemOpen
  \bibfield  {author} {\bibinfo {author} {\bibfnamefont {S.}~\bibnamefont
  {Kolkowitz}}, \bibinfo {author} {\bibfnamefont {S.~L.}\ \bibnamefont
  {Bromley}}, \bibinfo {author} {\bibfnamefont {T.}~\bibnamefont {Bothwell}},
  \bibinfo {author} {\bibfnamefont {M.~L.}\ \bibnamefont {Wall}}, \bibinfo
  {author} {\bibfnamefont {G.~E.}\ \bibnamefont {Marti}}, \bibinfo {author}
  {\bibfnamefont {A.~P.}\ \bibnamefont {Koller}}, \bibinfo {author}
  {\bibfnamefont {X.}~\bibnamefont {Zhang}}, \bibinfo {author} {\bibfnamefont
  {A.~M.}\ \bibnamefont {Rey}}, \ and\ \bibinfo {author} {\bibfnamefont
  {J.}~\bibnamefont {Ye}},\ }\href {\doibase 10.1038/nature20811} {\bibfield
  {journal} {\bibinfo  {journal} {Nature}\ }\textbf {\bibinfo {volume} {542}},\
  \bibinfo {pages} {66} (\bibinfo {year} {2017})}\BibitemShut {NoStop}%
\bibitem [{\citenamefont {Gadway}(2015)}]{Gadway-KSPACE}%
  \BibitemOpen
  \bibfield  {author} {\bibinfo {author} {\bibfnamefont {B.}~\bibnamefont
  {Gadway}},\ }\href {\doibase 10.1103/PhysRevA.92.043606} {\bibfield
  {journal} {\bibinfo  {journal} {Phys. Rev. A}\ }\textbf {\bibinfo {volume}
  {92}},\ \bibinfo {pages} {043606} (\bibinfo {year} {2015})}\BibitemShut
  {NoStop}%
\bibitem [{\citenamefont {Meier}\ \emph {et~al.}(2016)\citenamefont {Meier},
  \citenamefont {An},\ and\ \citenamefont {Gadway}}]{Meier-AtomOptics}%
  \BibitemOpen
  \bibfield  {author} {\bibinfo {author} {\bibfnamefont {E.~J.}\ \bibnamefont
  {Meier}}, \bibinfo {author} {\bibfnamefont {F.~A.}\ \bibnamefont {An}}, \
  and\ \bibinfo {author} {\bibfnamefont {B.}~\bibnamefont {Gadway}},\ }\href
  {\doibase 10.1103/PhysRevA.93.051602} {\bibfield  {journal} {\bibinfo
  {journal} {Phys. Rev. A}\ }\textbf {\bibinfo {volume} {93}},\ \bibinfo
  {pages} {051602} (\bibinfo {year} {2016})}\BibitemShut {NoStop}%
\bibitem [{\citenamefont {Chen}\ \emph {et~al.}(2021)\citenamefont {Chen},
  \citenamefont {Gou}, \citenamefont {Xie}, \citenamefont {Xiao}, \citenamefont
  {Yi}, \citenamefont {Jing},\ and\ \citenamefont {Yan}}]{Chen2021}%
  \BibitemOpen
  \bibfield  {author} {\bibinfo {author} {\bibfnamefont {T.}~\bibnamefont
  {Chen}}, \bibinfo {author} {\bibfnamefont {W.}~\bibnamefont {Gou}}, \bibinfo
  {author} {\bibfnamefont {D.}~\bibnamefont {Xie}}, \bibinfo {author}
  {\bibfnamefont {T.}~\bibnamefont {Xiao}}, \bibinfo {author} {\bibfnamefont
  {W.}~\bibnamefont {Yi}}, \bibinfo {author} {\bibfnamefont {J.}~\bibnamefont
  {Jing}}, \ and\ \bibinfo {author} {\bibfnamefont {B.}~\bibnamefont {Yan}},\
  }\href {\doibase 10.1038/s41534-021-00417-y} {\bibfield  {journal} {\bibinfo
  {journal} {npj Quantum Information}\ }\textbf {\bibinfo {volume} {7}},\
  \bibinfo {pages} {78} (\bibinfo {year} {2021})}\BibitemShut {NoStop}%
\bibitem [{\citenamefont {Price}\ \emph {et~al.}(2017)\citenamefont {Price},
  \citenamefont {Ozawa},\ and\ \citenamefont {Goldman}}]{Price-Shaking}%
  \BibitemOpen
  \bibfield  {author} {\bibinfo {author} {\bibfnamefont {H.~M.}\ \bibnamefont
  {Price}}, \bibinfo {author} {\bibfnamefont {T.}~\bibnamefont {Ozawa}}, \ and\
  \bibinfo {author} {\bibfnamefont {N.}~\bibnamefont {Goldman}},\ }\href
  {\doibase 10.1103/PhysRevA.95.023607} {\bibfield  {journal} {\bibinfo
  {journal} {Phys. Rev. A}\ }\textbf {\bibinfo {volume} {95}},\ \bibinfo
  {pages} {023607} (\bibinfo {year} {2017})}\BibitemShut {NoStop}%
\bibitem [{\citenamefont {Oliver}\ \emph {et~al.}(2021)\citenamefont {Oliver},
  \citenamefont {Smith}, \citenamefont {Easton}, \citenamefont {Salerno},
  \citenamefont {Guarrera}, \citenamefont {Goldman}, \citenamefont
  {Barontini},\ and\ \citenamefont {Price}}]{Barontini-Bloch}%
  \BibitemOpen
  \bibfield  {author} {\bibinfo {author} {\bibfnamefont {C.}~\bibnamefont
  {Oliver}}, \bibinfo {author} {\bibfnamefont {A.}~\bibnamefont {Smith}},
  \bibinfo {author} {\bibfnamefont {T.}~\bibnamefont {Easton}}, \bibinfo
  {author} {\bibfnamefont {G.}~\bibnamefont {Salerno}}, \bibinfo {author}
  {\bibfnamefont {V.}~\bibnamefont {Guarrera}}, \bibinfo {author}
  {\bibfnamefont {N.}~\bibnamefont {Goldman}}, \bibinfo {author} {\bibfnamefont
  {G.}~\bibnamefont {Barontini}}, \ and\ \bibinfo {author} {\bibfnamefont
  {H.~M.}\ \bibnamefont {Price}},\ }\href@noop {} {} (\bibinfo {year} {2021}),\
  \Eprint {http://arxiv.org/abs/2112.10648} {arXiv:2112.10648
  [cond-mat.quant-gas]} \BibitemShut {NoStop}%
\bibitem [{\citenamefont {Yuan}\ \emph {et~al.}(2021)\citenamefont {Yuan},
  \citenamefont {Dutt},\ and\ \citenamefont {Fan}}]{synth-freq}%
  \BibitemOpen
  \bibfield  {author} {\bibinfo {author} {\bibfnamefont {L.}~\bibnamefont
  {Yuan}}, \bibinfo {author} {\bibfnamefont {A.}~\bibnamefont {Dutt}}, \ and\
  \bibinfo {author} {\bibfnamefont {S.}~\bibnamefont {Fan}},\ }\href {\doibase
  10.1063/5.0056359} {\bibfield  {journal} {\bibinfo  {journal} {APL
  Photonics}\ }\textbf {\bibinfo {volume} {6}},\ \bibinfo {pages} {071102}
  (\bibinfo {year} {2021})}\BibitemShut {NoStop}%
\bibitem [{\citenamefont {Cardano}\ \emph {et~al.}(2017)\citenamefont
  {Cardano}, \citenamefont {D'Errico}, \citenamefont {Dauphin}, \citenamefont
  {Maffei}, \citenamefont {Piccirillo}, \citenamefont {de~Lisio}, \citenamefont
  {De~Filippis}, \citenamefont {Cataudella}, \citenamefont {Santamato},
  \citenamefont {Marrucci}, \citenamefont {Lewenstein},\ and\ \citenamefont
  {Massignan}}]{Cardano2017}%
  \BibitemOpen
  \bibfield  {author} {\bibinfo {author} {\bibfnamefont {F.}~\bibnamefont
  {Cardano}}, \bibinfo {author} {\bibfnamefont {A.}~\bibnamefont {D'Errico}},
  \bibinfo {author} {\bibfnamefont {A.}~\bibnamefont {Dauphin}}, \bibinfo
  {author} {\bibfnamefont {M.}~\bibnamefont {Maffei}}, \bibinfo {author}
  {\bibfnamefont {B.}~\bibnamefont {Piccirillo}}, \bibinfo {author}
  {\bibfnamefont {C.}~\bibnamefont {de~Lisio}}, \bibinfo {author}
  {\bibfnamefont {G.}~\bibnamefont {De~Filippis}}, \bibinfo {author}
  {\bibfnamefont {V.}~\bibnamefont {Cataudella}}, \bibinfo {author}
  {\bibfnamefont {E.}~\bibnamefont {Santamato}}, \bibinfo {author}
  {\bibfnamefont {L.}~\bibnamefont {Marrucci}}, \bibinfo {author}
  {\bibfnamefont {M.}~\bibnamefont {Lewenstein}}, \ and\ \bibinfo {author}
  {\bibfnamefont {P.}~\bibnamefont {Massignan}},\ }\href {\doibase
  10.1038/ncomms15516} {\bibfield  {journal} {\bibinfo  {journal} {Nature
  Communications}\ }\textbf {\bibinfo {volume} {8}},\ \bibinfo {pages} {15516}
  (\bibinfo {year} {2017})}\BibitemShut {NoStop}%
\bibitem [{\citenamefont {Chalabi}\ \emph {et~al.}(2019)\citenamefont
  {Chalabi}, \citenamefont {Barik}, \citenamefont {Mittal}, \citenamefont
  {Murphy}, \citenamefont {Hafezi},\ and\ \citenamefont
  {Waks}}]{Hafezi-TimeBin}%
  \BibitemOpen
  \bibfield  {author} {\bibinfo {author} {\bibfnamefont {H.}~\bibnamefont
  {Chalabi}}, \bibinfo {author} {\bibfnamefont {S.}~\bibnamefont {Barik}},
  \bibinfo {author} {\bibfnamefont {S.}~\bibnamefont {Mittal}}, \bibinfo
  {author} {\bibfnamefont {T.~E.}\ \bibnamefont {Murphy}}, \bibinfo {author}
  {\bibfnamefont {M.}~\bibnamefont {Hafezi}}, \ and\ \bibinfo {author}
  {\bibfnamefont {E.}~\bibnamefont {Waks}},\ }\href {\doibase
  10.1103/PhysRevLett.123.150503} {\bibfield  {journal} {\bibinfo  {journal}
  {Phys. Rev. Lett.}\ }\textbf {\bibinfo {volume} {123}},\ \bibinfo {pages}
  {150503} (\bibinfo {year} {2019})}\BibitemShut {NoStop}%
\bibitem [{\citenamefont {An}\ \emph {et~al.}(2018)\citenamefont {An},
  \citenamefont {Meier}, \citenamefont {Ang'ong'a},\ and\ \citenamefont
  {Gadway}}]{an2018corr}%
  \BibitemOpen
  \bibfield  {author} {\bibinfo {author} {\bibfnamefont {F.~A.}\ \bibnamefont
  {An}}, \bibinfo {author} {\bibfnamefont {E.~J.}\ \bibnamefont {Meier}},
  \bibinfo {author} {\bibfnamefont {J.}~\bibnamefont {Ang'ong'a}}, \ and\
  \bibinfo {author} {\bibfnamefont {B.}~\bibnamefont {Gadway}},\ }\href
  {\doibase 10.1103/PhysRevLett.120.040407} {\bibfield  {journal} {\bibinfo
  {journal} {Phys. Rev. Lett.}\ }\textbf {\bibinfo {volume} {120}},\ \bibinfo
  {pages} {040407} (\bibinfo {year} {2018})}\BibitemShut {NoStop}%
\bibitem [{\citenamefont {Bromley}\ \emph {et~al.}(2018)\citenamefont
  {Bromley}, \citenamefont {Kolkowitz}, \citenamefont {Bothwell}, \citenamefont
  {Kedar}, \citenamefont {Safavi-Naini}, \citenamefont {Wall}, \citenamefont
  {Salomon}, \citenamefont {Rey},\ and\ \citenamefont {Ye}}]{Bromley2018}%
  \BibitemOpen
  \bibfield  {author} {\bibinfo {author} {\bibfnamefont {S.~L.}\ \bibnamefont
  {Bromley}}, \bibinfo {author} {\bibfnamefont {S.}~\bibnamefont {Kolkowitz}},
  \bibinfo {author} {\bibfnamefont {T.}~\bibnamefont {Bothwell}}, \bibinfo
  {author} {\bibfnamefont {D.}~\bibnamefont {Kedar}}, \bibinfo {author}
  {\bibfnamefont {A.}~\bibnamefont {Safavi-Naini}}, \bibinfo {author}
  {\bibfnamefont {M.~L.}\ \bibnamefont {Wall}}, \bibinfo {author}
  {\bibfnamefont {C.}~\bibnamefont {Salomon}}, \bibinfo {author} {\bibfnamefont
  {A.~M.}\ \bibnamefont {Rey}}, \ and\ \bibinfo {author} {\bibfnamefont
  {J.}~\bibnamefont {Ye}},\ }\href {\doibase 10.1038/s41567-017-0029-0}
  {\bibfield  {journal} {\bibinfo  {journal} {Nature Physics}\ }\textbf
  {\bibinfo {volume} {14}},\ \bibinfo {pages} {399} (\bibinfo {year}
  {2018})}\BibitemShut {NoStop}%
\bibitem [{\citenamefont {Xie}\ \emph {et~al.}(2020)\citenamefont {Xie},
  \citenamefont {Deng}, \citenamefont {Xiao}, \citenamefont {Gou},
  \citenamefont {Chen}, \citenamefont {Yi},\ and\ \citenamefont
  {Yan}}]{Xie-Walk}%
  \BibitemOpen
  \bibfield  {author} {\bibinfo {author} {\bibfnamefont {D.}~\bibnamefont
  {Xie}}, \bibinfo {author} {\bibfnamefont {T.-S.}\ \bibnamefont {Deng}},
  \bibinfo {author} {\bibfnamefont {T.}~\bibnamefont {Xiao}}, \bibinfo {author}
  {\bibfnamefont {W.}~\bibnamefont {Gou}}, \bibinfo {author} {\bibfnamefont
  {T.}~\bibnamefont {Chen}}, \bibinfo {author} {\bibfnamefont {W.}~\bibnamefont
  {Yi}}, \ and\ \bibinfo {author} {\bibfnamefont {B.}~\bibnamefont {Yan}},\
  }\href {\doibase 10.1103/PhysRevLett.124.050502} {\bibfield  {journal}
  {\bibinfo  {journal} {Phys. Rev. Lett.}\ }\textbf {\bibinfo {volume} {124}},\
  \bibinfo {pages} {050502} (\bibinfo {year} {2020})}\BibitemShut {NoStop}%
\bibitem [{\citenamefont {An}\ \emph {et~al.}(2021)\citenamefont {An},
  \citenamefont {Sundar}, \citenamefont {Hou}, \citenamefont {Luo},
  \citenamefont {Meier}, \citenamefont {Zhang}, \citenamefont {Hazzard},\ and\
  \citenamefont {Gadway}}]{An-nonlinear}%
  \BibitemOpen
  \bibfield  {author} {\bibinfo {author} {\bibfnamefont {F.~A.}\ \bibnamefont
  {An}}, \bibinfo {author} {\bibfnamefont {B.}~\bibnamefont {Sundar}}, \bibinfo
  {author} {\bibfnamefont {J.}~\bibnamefont {Hou}}, \bibinfo {author}
  {\bibfnamefont {X.-W.}\ \bibnamefont {Luo}}, \bibinfo {author} {\bibfnamefont
  {E.~J.}\ \bibnamefont {Meier}}, \bibinfo {author} {\bibfnamefont
  {C.}~\bibnamefont {Zhang}}, \bibinfo {author} {\bibfnamefont {K.~R.~A.}\
  \bibnamefont {Hazzard}}, \ and\ \bibinfo {author} {\bibfnamefont
  {B.}~\bibnamefont {Gadway}},\ }\href {\doibase
  10.1103/PhysRevLett.127.130401} {\bibfield  {journal} {\bibinfo  {journal}
  {Phys. Rev. Lett.}\ }\textbf {\bibinfo {volume} {127}},\ \bibinfo {pages}
  {130401} (\bibinfo {year} {2021})}\BibitemShut {NoStop}%
\bibitem [{\citenamefont {Wang}\ \emph {et~al.}(2022)\citenamefont {Wang},
  \citenamefont {Zhang}, \citenamefont {Li}, \citenamefont {Wu}, \citenamefont
  {Liu}, \citenamefont {Mei}, \citenamefont {Hu}, \citenamefont {Xiao},
  \citenamefont {Ma}, \citenamefont {Chin},\ and\ \citenamefont
  {Jia}}]{Wang-Interactions}%
  \BibitemOpen
  \bibfield  {author} {\bibinfo {author} {\bibfnamefont {Y.}~\bibnamefont
  {Wang}}, \bibinfo {author} {\bibfnamefont {J.-H.}\ \bibnamefont {Zhang}},
  \bibinfo {author} {\bibfnamefont {Y.}~\bibnamefont {Li}}, \bibinfo {author}
  {\bibfnamefont {J.}~\bibnamefont {Wu}}, \bibinfo {author} {\bibfnamefont
  {W.}~\bibnamefont {Liu}}, \bibinfo {author} {\bibfnamefont {F.}~\bibnamefont
  {Mei}}, \bibinfo {author} {\bibfnamefont {Y.}~\bibnamefont {Hu}}, \bibinfo
  {author} {\bibfnamefont {L.}~\bibnamefont {Xiao}}, \bibinfo {author}
  {\bibfnamefont {J.}~\bibnamefont {Ma}}, \bibinfo {author} {\bibfnamefont
  {C.}~\bibnamefont {Chin}}, \ and\ \bibinfo {author} {\bibfnamefont
  {S.}~\bibnamefont {Jia}},\ }\href {\doibase 10.1103/PhysRevLett.129.103401}
  {\bibfield  {journal} {\bibinfo  {journal} {Phys. Rev. Lett.}\ }\textbf
  {\bibinfo {volume} {129}},\ \bibinfo {pages} {103401} (\bibinfo {year}
  {2022})}\BibitemShut {NoStop}%
\bibitem [{\citenamefont {Wimmer}\ \emph {et~al.}(2021)\citenamefont {Wimmer},
  \citenamefont {Monika}, \citenamefont {Carusotto}, \citenamefont {Peschel},\
  and\ \citenamefont {Price}}]{Price-photons}%
  \BibitemOpen
  \bibfield  {author} {\bibinfo {author} {\bibfnamefont {M.}~\bibnamefont
  {Wimmer}}, \bibinfo {author} {\bibfnamefont {M.}~\bibnamefont {Monika}},
  \bibinfo {author} {\bibfnamefont {I.}~\bibnamefont {Carusotto}}, \bibinfo
  {author} {\bibfnamefont {U.}~\bibnamefont {Peschel}}, \ and\ \bibinfo
  {author} {\bibfnamefont {H.~M.}\ \bibnamefont {Price}},\ }\href {\doibase
  10.1103/PhysRevLett.127.163901} {\bibfield  {journal} {\bibinfo  {journal}
  {Phys. Rev. Lett.}\ }\textbf {\bibinfo {volume} {127}},\ \bibinfo {pages}
  {163901} (\bibinfo {year} {2021})}\BibitemShut {NoStop}%
\bibitem [{\citenamefont {Zhou}\ \emph {et~al.}(2022)\citenamefont {Zhou},
  \citenamefont {Cappellini}, \citenamefont {Tusi}, \citenamefont {Franchi},
  \citenamefont {Parravicini}, \citenamefont {Repellin}, \citenamefont
  {Greschner}, \citenamefont {Inguscio}, \citenamefont {Giamarchi},
  \citenamefont {Filippone}, \citenamefont {Catani},\ and\ \citenamefont
  {Fallani}}]{zhou2022observation}%
  \BibitemOpen
  \bibfield  {author} {\bibinfo {author} {\bibfnamefont {T.~W.}\ \bibnamefont
  {Zhou}}, \bibinfo {author} {\bibfnamefont {G.}~\bibnamefont {Cappellini}},
  \bibinfo {author} {\bibfnamefont {D.}~\bibnamefont {Tusi}}, \bibinfo {author}
  {\bibfnamefont {L.}~\bibnamefont {Franchi}}, \bibinfo {author} {\bibfnamefont
  {J.}~\bibnamefont {Parravicini}}, \bibinfo {author} {\bibfnamefont
  {C.}~\bibnamefont {Repellin}}, \bibinfo {author} {\bibfnamefont
  {S.}~\bibnamefont {Greschner}}, \bibinfo {author} {\bibfnamefont
  {M.}~\bibnamefont {Inguscio}}, \bibinfo {author} {\bibfnamefont
  {T.}~\bibnamefont {Giamarchi}}, \bibinfo {author} {\bibfnamefont
  {M.}~\bibnamefont {Filippone}}, \bibinfo {author} {\bibfnamefont
  {J.}~\bibnamefont {Catani}}, \ and\ \bibinfo {author} {\bibfnamefont
  {L.}~\bibnamefont {Fallani}},\ }\href@noop {} {} (\bibinfo {year} {2022}),\
  \Eprint {http://arxiv.org/abs/2205.13567} {arXiv:2205.13567} \BibitemShut
  {NoStop}%
\bibitem [{\citenamefont {Sundar}\ \emph {et~al.}(2018)\citenamefont {Sundar},
  \citenamefont {Gadway},\ and\ \citenamefont {Hazzard}}]{Sundar2018}%
  \BibitemOpen
  \bibfield  {author} {\bibinfo {author} {\bibfnamefont {B.}~\bibnamefont
  {Sundar}}, \bibinfo {author} {\bibfnamefont {B.}~\bibnamefont {Gadway}}, \
  and\ \bibinfo {author} {\bibfnamefont {K.~R.~A.}\ \bibnamefont {Hazzard}},\
  }\href {\doibase 10.1038/s41598-018-21699-x} {\bibfield  {journal} {\bibinfo
  {journal} {Scientific Reports}\ }\textbf {\bibinfo {volume} {8}},\ \bibinfo
  {pages} {3422} (\bibinfo {year} {2018})}\BibitemShut {NoStop}%
\bibitem [{\citenamefont {Sundar}\ \emph {et~al.}(2019)\citenamefont {Sundar},
  \citenamefont {Thibodeau}, \citenamefont {Wang}, \citenamefont {Gadway},\
  and\ \citenamefont {Hazzard}}]{SundarPRA}%
  \BibitemOpen
  \bibfield  {author} {\bibinfo {author} {\bibfnamefont {B.}~\bibnamefont
  {Sundar}}, \bibinfo {author} {\bibfnamefont {M.}~\bibnamefont {Thibodeau}},
  \bibinfo {author} {\bibfnamefont {Z.}~\bibnamefont {Wang}}, \bibinfo {author}
  {\bibfnamefont {B.}~\bibnamefont {Gadway}}, \ and\ \bibinfo {author}
  {\bibfnamefont {K.~R.~A.}\ \bibnamefont {Hazzard}},\ }\href {\doibase
  10.1103/PhysRevA.99.013624} {\bibfield  {journal} {\bibinfo  {journal} {Phys.
  Rev. A}\ }\textbf {\bibinfo {volume} {99}},\ \bibinfo {pages} {013624}
  (\bibinfo {year} {2019})}\BibitemShut {NoStop}%
\bibitem [{\citenamefont {Feng}\ \emph {et~al.}(2022)\citenamefont {Feng},
  \citenamefont {Manetsch}, \citenamefont {Rousseau}, \citenamefont {Hazzard},\
  and\ \citenamefont {Scalettar}}]{Membranes}%
  \BibitemOpen
  \bibfield  {author} {\bibinfo {author} {\bibfnamefont {C.}~\bibnamefont
  {Feng}}, \bibinfo {author} {\bibfnamefont {H.}~\bibnamefont {Manetsch}},
  \bibinfo {author} {\bibfnamefont {V.~G.}\ \bibnamefont {Rousseau}}, \bibinfo
  {author} {\bibfnamefont {K.~R.~A.}\ \bibnamefont {Hazzard}}, \ and\ \bibinfo
  {author} {\bibfnamefont {R.}~\bibnamefont {Scalettar}},\ }\href {\doibase
  10.1103/PhysRevA.105.063320} {\bibfield  {journal} {\bibinfo  {journal}
  {Phys. Rev. A}\ }\textbf {\bibinfo {volume} {105}},\ \bibinfo {pages}
  {063320} (\bibinfo {year} {2022})}\BibitemShut {NoStop}%
\bibitem [{\citenamefont {Yan}\ \emph {et~al.}(2013)\citenamefont {Yan},
  \citenamefont {Moses}, \citenamefont {Gadway}, \citenamefont {Covey},
  \citenamefont {Hazzard}, \citenamefont {Rey}, \citenamefont {Jin},\ and\
  \citenamefont {Ye}}]{Yan2013}%
  \BibitemOpen
  \bibfield  {author} {\bibinfo {author} {\bibfnamefont {B.}~\bibnamefont
  {Yan}}, \bibinfo {author} {\bibfnamefont {S.~A.}\ \bibnamefont {Moses}},
  \bibinfo {author} {\bibfnamefont {B.}~\bibnamefont {Gadway}}, \bibinfo
  {author} {\bibfnamefont {J.~P.}\ \bibnamefont {Covey}}, \bibinfo {author}
  {\bibfnamefont {K.~R.~A.}\ \bibnamefont {Hazzard}}, \bibinfo {author}
  {\bibfnamefont {A.~M.}\ \bibnamefont {Rey}}, \bibinfo {author} {\bibfnamefont
  {D.~S.}\ \bibnamefont {Jin}}, \ and\ \bibinfo {author} {\bibfnamefont
  {J.}~\bibnamefont {Ye}},\ }\href {\doibase 10.1038/nature12483} {\bibfield
  {journal} {\bibinfo  {journal} {Nature}\ }\textbf {\bibinfo {volume} {501}},\
  \bibinfo {pages} {521} (\bibinfo {year} {2013})}\BibitemShut {NoStop}%
\bibitem [{\citenamefont {Browaeys}\ \emph {et~al.}(2016)\citenamefont
  {Browaeys}, \citenamefont {Barredo},\ and\ \citenamefont
  {Lahaye}}]{Browaeys_2016}%
  \BibitemOpen
  \bibfield  {author} {\bibinfo {author} {\bibfnamefont {A.}~\bibnamefont
  {Browaeys}}, \bibinfo {author} {\bibfnamefont {D.}~\bibnamefont {Barredo}}, \
  and\ \bibinfo {author} {\bibfnamefont {T.}~\bibnamefont {Lahaye}},\ }\href
  {\doibase 10.1088/0953-4075/49/15/152001} {\bibfield  {journal} {\bibinfo
  {journal} {Journal of Physics B: Atomic, Molecular and Optical Physics}\
  }\textbf {\bibinfo {volume} {49}},\ \bibinfo {pages} {152001} (\bibinfo
  {year} {2016})}\BibitemShut {NoStop}%
\bibitem [{\citenamefont {Gadway}\ and\ \citenamefont
  {Yan}(2016)}]{Gadway_2016}%
  \BibitemOpen
  \bibfield  {author} {\bibinfo {author} {\bibfnamefont {B.}~\bibnamefont
  {Gadway}}\ and\ \bibinfo {author} {\bibfnamefont {B.}~\bibnamefont {Yan}},\
  }\href {\doibase 10.1088/0953-4075/49/15/152002} {\bibfield  {journal}
  {\bibinfo  {journal} {Journal of Physics B: Atomic, Molecular and Optical
  Physics}\ }\textbf {\bibinfo {volume} {49}},\ \bibinfo {pages} {152002}
  (\bibinfo {year} {2016})}\BibitemShut {NoStop}%
\bibitem [{\citenamefont {Kanungo}\ \emph {et~al.}(2022)\citenamefont
  {Kanungo}, \citenamefont {Whalen}, \citenamefont {Lu}, \citenamefont {Yuan},
  \citenamefont {Dasgupta}, \citenamefont {Dunning}, \citenamefont {Hazzard},\
  and\ \citenamefont {Killian}}]{Kanungo2022}%
  \BibitemOpen
  \bibfield  {author} {\bibinfo {author} {\bibfnamefont {S.~K.}\ \bibnamefont
  {Kanungo}}, \bibinfo {author} {\bibfnamefont {J.~D.}\ \bibnamefont {Whalen}},
  \bibinfo {author} {\bibfnamefont {Y.}~\bibnamefont {Lu}}, \bibinfo {author}
  {\bibfnamefont {M.}~\bibnamefont {Yuan}}, \bibinfo {author} {\bibfnamefont
  {S.}~\bibnamefont {Dasgupta}}, \bibinfo {author} {\bibfnamefont {F.~B.}\
  \bibnamefont {Dunning}}, \bibinfo {author} {\bibfnamefont {K.~R.~A.}\
  \bibnamefont {Hazzard}}, \ and\ \bibinfo {author} {\bibfnamefont {T.~C.}\
  \bibnamefont {Killian}},\ }\href {\doibase 10.1038/s41467-022-28550-y}
  {\bibfield  {journal} {\bibinfo  {journal} {Nature Communications}\ }\textbf
  {\bibinfo {volume} {13}},\ \bibinfo {pages} {972} (\bibinfo {year}
  {2022})}\BibitemShut {NoStop}%
\bibitem [{Note1()}]{Note1}%
  \BibitemOpen
  \bibinfo {note} {See also Ref.~\cite {Blackmore} for steps towards molecular
  synthetic dimensions, as well as related early work in Rydbergs and
  molecules~\cite {Signoles2014,floss2015observation}}\BibitemShut {NoStop}%
\bibitem [{Sup()}]{SuppMats}%
  \BibitemOpen
  \href@noop {} {}\bibinfo {note} {See Supplementary Material, and references
  contained therein~\cite{Tuchendler08,Walker2012,sibalic17soft}, for more
  experimental details on the synthetic lattice calibration and dipole-dipole
  interactions.}\BibitemShut {Stop}%
\bibitem [{\citenamefont {An}\ \emph {et~al.}(2017)\citenamefont {An},
  \citenamefont {Meier},\ and\ \citenamefont {Gadway}}]{An-FluxLadder}%
  \BibitemOpen
  \bibfield  {author} {\bibinfo {author} {\bibfnamefont {F.~A.}\ \bibnamefont
  {An}}, \bibinfo {author} {\bibfnamefont {E.~J.}\ \bibnamefont {Meier}}, \
  and\ \bibinfo {author} {\bibfnamefont {B.}~\bibnamefont {Gadway}},\ }\href
  {\doibase 10.1126/sciadv.1602685} {\bibfield  {journal} {\bibinfo  {journal}
  {Sci. Adv.}\ }\textbf {\bibinfo {volume} {3}},\ \bibinfo {pages} {e1602685}
  (\bibinfo {year} {2017})}\BibitemShut {NoStop}%
\bibitem [{\citenamefont {Gou}\ \emph {et~al.}(2020)\citenamefont {Gou},
  \citenamefont {Chen}, \citenamefont {Xie}, \citenamefont {Xiao},
  \citenamefont {Deng}, \citenamefont {Gadway}, \citenamefont {Yi},\ and\
  \citenamefont {Yan}}]{FluxNonrecip}%
  \BibitemOpen
  \bibfield  {author} {\bibinfo {author} {\bibfnamefont {W.}~\bibnamefont
  {Gou}}, \bibinfo {author} {\bibfnamefont {T.}~\bibnamefont {Chen}}, \bibinfo
  {author} {\bibfnamefont {D.}~\bibnamefont {Xie}}, \bibinfo {author}
  {\bibfnamefont {T.}~\bibnamefont {Xiao}}, \bibinfo {author} {\bibfnamefont
  {T.-S.}\ \bibnamefont {Deng}}, \bibinfo {author} {\bibfnamefont
  {B.}~\bibnamefont {Gadway}}, \bibinfo {author} {\bibfnamefont
  {W.}~\bibnamefont {Yi}}, \ and\ \bibinfo {author} {\bibfnamefont
  {B.}~\bibnamefont {Yan}},\ }\href {\doibase 10.1103/PhysRevLett.124.070402}
  {\bibfield  {journal} {\bibinfo  {journal} {Phys. Rev. Lett.}\ }\textbf
  {\bibinfo {volume} {124}},\ \bibinfo {pages} {070402} (\bibinfo {year}
  {2020})}\BibitemShut {NoStop}%
\bibitem [{\citenamefont {Shen}\ \emph {et~al.}(2022)\citenamefont {Shen},
  \citenamefont {Luo}, \citenamefont {Huang}, \citenamefont {Clark},
  \citenamefont {El-Khadra}, \citenamefont {Gadway},\ and\ \citenamefont
  {Draper}}]{tHooft}%
  \BibitemOpen
  \bibfield  {author} {\bibinfo {author} {\bibfnamefont {J.}~\bibnamefont
  {Shen}}, \bibinfo {author} {\bibfnamefont {D.}~\bibnamefont {Luo}}, \bibinfo
  {author} {\bibfnamefont {C.}~\bibnamefont {Huang}}, \bibinfo {author}
  {\bibfnamefont {B.~K.}\ \bibnamefont {Clark}}, \bibinfo {author}
  {\bibfnamefont {A.~X.}\ \bibnamefont {El-Khadra}}, \bibinfo {author}
  {\bibfnamefont {B.}~\bibnamefont {Gadway}}, \ and\ \bibinfo {author}
  {\bibfnamefont {P.}~\bibnamefont {Draper}},\ }\href {\doibase
  10.1103/PhysRevD.105.074505} {\bibfield  {journal} {\bibinfo  {journal}
  {Phys. Rev. D}\ }\textbf {\bibinfo {volume} {105}},\ \bibinfo {pages}
  {074505} (\bibinfo {year} {2022})}\BibitemShut {NoStop}%
\bibitem [{\citenamefont {Liang}\ \emph {et~al.}(2021)\citenamefont {Liang},
  \citenamefont {Trypogeorgos}, \citenamefont {Vald\'es-Curiel}, \citenamefont
  {Tao}, \citenamefont {Zhao},\ and\ \citenamefont
  {Spielman}}]{Liang-Harp-Hof}%
  \BibitemOpen
  \bibfield  {author} {\bibinfo {author} {\bibfnamefont {Q.-Y.}\ \bibnamefont
  {Liang}}, \bibinfo {author} {\bibfnamefont {D.}~\bibnamefont {Trypogeorgos}},
  \bibinfo {author} {\bibfnamefont {A.}~\bibnamefont {Vald\'es-Curiel}},
  \bibinfo {author} {\bibfnamefont {J.}~\bibnamefont {Tao}}, \bibinfo {author}
  {\bibfnamefont {M.}~\bibnamefont {Zhao}}, \ and\ \bibinfo {author}
  {\bibfnamefont {I.~B.}\ \bibnamefont {Spielman}},\ }\href {\doibase
  10.1103/PhysRevResearch.3.023058} {\bibfield  {journal} {\bibinfo  {journal}
  {Phys. Rev. Res.}\ }\textbf {\bibinfo {volume} {3}},\ \bibinfo {pages}
  {023058} (\bibinfo {year} {2021})}\BibitemShut {NoStop}%
\bibitem [{\citenamefont {Fabre}\ \emph {et~al.}(2022)\citenamefont {Fabre},
  \citenamefont {Bouhiron}, \citenamefont {Satoor}, \citenamefont {Lopes},\
  and\ \citenamefont {Nascimbene}}]{Nasc-Laughlin}%
  \BibitemOpen
  \bibfield  {author} {\bibinfo {author} {\bibfnamefont {A.}~\bibnamefont
  {Fabre}}, \bibinfo {author} {\bibfnamefont {J.-B.}\ \bibnamefont {Bouhiron}},
  \bibinfo {author} {\bibfnamefont {T.}~\bibnamefont {Satoor}}, \bibinfo
  {author} {\bibfnamefont {R.}~\bibnamefont {Lopes}}, \ and\ \bibinfo {author}
  {\bibfnamefont {S.}~\bibnamefont {Nascimbene}},\ }\href {\doibase
  10.1103/PhysRevLett.128.173202} {\bibfield  {journal} {\bibinfo  {journal}
  {Phys. Rev. Lett.}\ }\textbf {\bibinfo {volume} {128}},\ \bibinfo {pages}
  {173202} (\bibinfo {year} {2022})}\BibitemShut {NoStop}%
\bibitem [{\citenamefont {Li}\ \emph {et~al.}(2022)\citenamefont {Li},
  \citenamefont {Yan}, \citenamefont {Feng}, \citenamefont {Choudhury},
  \citenamefont {Blasing}, \citenamefont {Zhou},\ and\ \citenamefont
  {Chen}}]{Yong-Flux-Cylinder}%
  \BibitemOpen
  \bibfield  {author} {\bibinfo {author} {\bibfnamefont {C.-H.}\ \bibnamefont
  {Li}}, \bibinfo {author} {\bibfnamefont {Y.}~\bibnamefont {Yan}}, \bibinfo
  {author} {\bibfnamefont {S.-W.}\ \bibnamefont {Feng}}, \bibinfo {author}
  {\bibfnamefont {S.}~\bibnamefont {Choudhury}}, \bibinfo {author}
  {\bibfnamefont {D.~B.}\ \bibnamefont {Blasing}}, \bibinfo {author}
  {\bibfnamefont {Q.}~\bibnamefont {Zhou}}, \ and\ \bibinfo {author}
  {\bibfnamefont {Y.~P.}\ \bibnamefont {Chen}},\ }\href {\doibase
  10.1103/PRXQuantum.3.010316} {\bibfield  {journal} {\bibinfo  {journal} {PRX
  Quantum}\ }\textbf {\bibinfo {volume} {3}},\ \bibinfo {pages} {010316}
  (\bibinfo {year} {2022})}\BibitemShut {NoStop}%
\bibitem [{\citenamefont {Browaeys}\ and\ \citenamefont
  {Lahaye}(2020)}]{Browaeys2020-NatPhys}%
  \BibitemOpen
  \bibfield  {author} {\bibinfo {author} {\bibfnamefont {A.}~\bibnamefont
  {Browaeys}}\ and\ \bibinfo {author} {\bibfnamefont {T.}~\bibnamefont
  {Lahaye}},\ }\href {\doibase 10.1038/s41567-019-0733-z} {\bibfield  {journal}
  {\bibinfo  {journal} {Nature Physics}\ }\textbf {\bibinfo {volume} {16}},\
  \bibinfo {pages} {132} (\bibinfo {year} {2020})}\BibitemShut {NoStop}%
\bibitem [{\citenamefont {Kaufman}\ and\ \citenamefont
  {Ni}(2021)}]{Kaufman2021}%
  \BibitemOpen
  \bibfield  {author} {\bibinfo {author} {\bibfnamefont {A.~M.}\ \bibnamefont
  {Kaufman}}\ and\ \bibinfo {author} {\bibfnamefont {K.-K.}\ \bibnamefont
  {Ni}},\ }\href {\doibase 10.1038/s41567-021-01357-2} {\bibfield  {journal}
  {\bibinfo  {journal} {Nature Physics}\ }\textbf {\bibinfo {volume} {17}},\
  \bibinfo {pages} {1324} (\bibinfo {year} {2021})}\BibitemShut {NoStop}%
\bibitem [{\citenamefont {Lorenz}\ \emph {et~al.}(2021)\citenamefont {Lorenz},
  \citenamefont {Festa}, \citenamefont {Steinert},\ and\ \citenamefont
  {Gross}}]{Gross-SciPost}%
  \BibitemOpen
  \bibfield  {author} {\bibinfo {author} {\bibfnamefont {N.}~\bibnamefont
  {Lorenz}}, \bibinfo {author} {\bibfnamefont {L.}~\bibnamefont {Festa}},
  \bibinfo {author} {\bibfnamefont {L.-M.}\ \bibnamefont {Steinert}}, \ and\
  \bibinfo {author} {\bibfnamefont {C.}~\bibnamefont {Gross}},\ }\href
  {\doibase 10.21468/SciPostPhys.10.3.052} {\bibfield  {journal} {\bibinfo
  {journal} {SciPost Phys.}\ }\textbf {\bibinfo {volume} {10}},\ \bibinfo
  {pages} {052} (\bibinfo {year} {2021})}\BibitemShut {NoStop}%
\bibitem [{\citenamefont {Ang'ong'a}\ \emph {et~al.}(2022)\citenamefont
  {Ang'ong'a}, \citenamefont {Huang}, \citenamefont {Covey},\ and\
  \citenamefont {Gadway}}]{JacksonPRR}%
  \BibitemOpen
  \bibfield  {author} {\bibinfo {author} {\bibfnamefont {J.}~\bibnamefont
  {Ang'ong'a}}, \bibinfo {author} {\bibfnamefont {C.}~\bibnamefont {Huang}},
  \bibinfo {author} {\bibfnamefont {J.~P.}\ \bibnamefont {Covey}}, \ and\
  \bibinfo {author} {\bibfnamefont {B.}~\bibnamefont {Gadway}},\ }\href
  {\doibase 10.1103/PhysRevResearch.4.013240} {\bibfield  {journal} {\bibinfo
  {journal} {Phys. Rev. Res.}\ }\textbf {\bibinfo {volume} {4}},\ \bibinfo
  {pages} {013240} (\bibinfo {year} {2022})}\BibitemShut {NoStop}%
\bibitem [{\citenamefont {Salomon}\ \emph {et~al.}(2013)\citenamefont
  {Salomon}, \citenamefont {Fouch{\'{e}}}, \citenamefont {Wang}, \citenamefont
  {Aspect}, \citenamefont {Bouyer},\ and\ \citenamefont
  {Bourdel}}]{Salomon_2013}%
  \BibitemOpen
  \bibfield  {author} {\bibinfo {author} {\bibfnamefont {G.}~\bibnamefont
  {Salomon}}, \bibinfo {author} {\bibfnamefont {L.}~\bibnamefont
  {Fouch{\'{e}}}}, \bibinfo {author} {\bibfnamefont {P.}~\bibnamefont {Wang}},
  \bibinfo {author} {\bibfnamefont {A.}~\bibnamefont {Aspect}}, \bibinfo
  {author} {\bibfnamefont {P.}~\bibnamefont {Bouyer}}, \ and\ \bibinfo {author}
  {\bibfnamefont {T.}~\bibnamefont {Bourdel}},\ }\href {\doibase
  https://doi.org/10.1209/0295-5075/104/63002} {\bibfield  {journal} {\bibinfo
  {journal} {{EPL} (Europhysics Letters)}\ }\textbf {\bibinfo {volume} {104}},\
  \bibinfo {pages} {63002} (\bibinfo {year} {2013})}\BibitemShut {NoStop}%
\bibitem [{\citenamefont {Cubel}\ \emph {et~al.}(2005)\citenamefont {Cubel},
  \citenamefont {Teo}, \citenamefont {Malinovsky}, \citenamefont {Guest},
  \citenamefont {Reinhard}, \citenamefont {Knuffman}, \citenamefont {Berman},\
  and\ \citenamefont {Raithel}}]{Cubel-STIRAP}%
  \BibitemOpen
  \bibfield  {author} {\bibinfo {author} {\bibfnamefont {T.}~\bibnamefont
  {Cubel}}, \bibinfo {author} {\bibfnamefont {B.~K.}\ \bibnamefont {Teo}},
  \bibinfo {author} {\bibfnamefont {V.~S.}\ \bibnamefont {Malinovsky}},
  \bibinfo {author} {\bibfnamefont {J.~R.}\ \bibnamefont {Guest}}, \bibinfo
  {author} {\bibfnamefont {A.}~\bibnamefont {Reinhard}}, \bibinfo {author}
  {\bibfnamefont {B.}~\bibnamefont {Knuffman}}, \bibinfo {author}
  {\bibfnamefont {P.~R.}\ \bibnamefont {Berman}}, \ and\ \bibinfo {author}
  {\bibfnamefont {G.}~\bibnamefont {Raithel}},\ }\href {\doibase
  10.1103/PhysRevA.72.023405} {\bibfield  {journal} {\bibinfo  {journal} {Phys.
  Rev. A}\ }\textbf {\bibinfo {volume} {72}},\ \bibinfo {pages} {023405}
  (\bibinfo {year} {2005})}\BibitemShut {NoStop}%
\bibitem [{\citenamefont {de~Léséleuc}\ \emph {et~al.}(2019)\citenamefont
  {de~Léséleuc}, \citenamefont {Lienhard}, \citenamefont {Scholl},
  \citenamefont {Barredo}, \citenamefont {Weber}, \citenamefont {Lang},
  \citenamefont {Büchler}, \citenamefont {Lahaye},\ and\ \citenamefont
  {Browaeys}}]{sylvain18}%
  \BibitemOpen
  \bibfield  {author} {\bibinfo {author} {\bibfnamefont {S.}~\bibnamefont
  {de~Léséleuc}}, \bibinfo {author} {\bibfnamefont {V.}~\bibnamefont
  {Lienhard}}, \bibinfo {author} {\bibfnamefont {P.}~\bibnamefont {Scholl}},
  \bibinfo {author} {\bibfnamefont {D.}~\bibnamefont {Barredo}}, \bibinfo
  {author} {\bibfnamefont {S.}~\bibnamefont {Weber}}, \bibinfo {author}
  {\bibfnamefont {N.}~\bibnamefont {Lang}}, \bibinfo {author} {\bibfnamefont
  {H.~P.}\ \bibnamefont {Büchler}}, \bibinfo {author} {\bibfnamefont
  {T.}~\bibnamefont {Lahaye}}, \ and\ \bibinfo {author} {\bibfnamefont
  {A.}~\bibnamefont {Browaeys}},\ }\href {\doibase 10.1126/science.aav9105}
  {\bibfield  {journal} {\bibinfo  {journal} {Science}\ }\textbf {\bibinfo
  {volume} {365}},\ \bibinfo {pages} {775} (\bibinfo {year}
  {2019})}\BibitemShut {NoStop}%
\bibitem [{\citenamefont {de~L\'es\'eleuc}\ \emph {et~al.}(2017)\citenamefont
  {de~L\'es\'eleuc}, \citenamefont {Barredo}, \citenamefont {Lienhard},
  \citenamefont {Browaeys},\ and\ \citenamefont {Lahaye}}]{Sylvain}%
  \BibitemOpen
  \bibfield  {author} {\bibinfo {author} {\bibfnamefont {S.}~\bibnamefont
  {de~L\'es\'eleuc}}, \bibinfo {author} {\bibfnamefont {D.}~\bibnamefont
  {Barredo}}, \bibinfo {author} {\bibfnamefont {V.}~\bibnamefont {Lienhard}},
  \bibinfo {author} {\bibfnamefont {A.}~\bibnamefont {Browaeys}}, \ and\
  \bibinfo {author} {\bibfnamefont {T.}~\bibnamefont {Lahaye}},\ }\href
  {\doibase 10.1103/PhysRevLett.119.053202} {\bibfield  {journal} {\bibinfo
  {journal} {Phys. Rev. Lett.}\ }\textbf {\bibinfo {volume} {119}},\ \bibinfo
  {pages} {053202} (\bibinfo {year} {2017})}\BibitemShut {NoStop}%
\bibitem [{\citenamefont {Preiss}\ \emph {et~al.}(2015)\citenamefont {Preiss},
  \citenamefont {Ma}, \citenamefont {Tai}, \citenamefont {Lukin}, \citenamefont
  {Rispoli}, \citenamefont {Zupancic}, \citenamefont {Lahini}, \citenamefont
  {Islam},\ and\ \citenamefont {Greiner}}]{Preiss-QWalk}%
  \BibitemOpen
  \bibfield  {author} {\bibinfo {author} {\bibfnamefont {P.~M.}\ \bibnamefont
  {Preiss}}, \bibinfo {author} {\bibfnamefont {R.}~\bibnamefont {Ma}}, \bibinfo
  {author} {\bibfnamefont {M.~E.}\ \bibnamefont {Tai}}, \bibinfo {author}
  {\bibfnamefont {A.}~\bibnamefont {Lukin}}, \bibinfo {author} {\bibfnamefont
  {M.}~\bibnamefont {Rispoli}}, \bibinfo {author} {\bibfnamefont
  {P.}~\bibnamefont {Zupancic}}, \bibinfo {author} {\bibfnamefont
  {Y.}~\bibnamefont {Lahini}}, \bibinfo {author} {\bibfnamefont
  {R.}~\bibnamefont {Islam}}, \ and\ \bibinfo {author} {\bibfnamefont
  {M.}~\bibnamefont {Greiner}},\ }\href {\doibase 10.1126/science.1260364}
  {\bibfield  {journal} {\bibinfo  {journal} {Science}\ }\textbf {\bibinfo
  {volume} {347}},\ \bibinfo {pages} {1229} (\bibinfo {year}
  {2015})}\BibitemShut {NoStop}%
\bibitem [{\citenamefont {Sala}\ \emph {et~al.}(2020)\citenamefont {Sala},
  \citenamefont {Rakovszky}, \citenamefont {Verresen}, \citenamefont {Knap},\
  and\ \citenamefont {Pollmann}}]{Frag1}%
  \BibitemOpen
  \bibfield  {author} {\bibinfo {author} {\bibfnamefont {P.}~\bibnamefont
  {Sala}}, \bibinfo {author} {\bibfnamefont {T.}~\bibnamefont {Rakovszky}},
  \bibinfo {author} {\bibfnamefont {R.}~\bibnamefont {Verresen}}, \bibinfo
  {author} {\bibfnamefont {M.}~\bibnamefont {Knap}}, \ and\ \bibinfo {author}
  {\bibfnamefont {F.}~\bibnamefont {Pollmann}},\ }\href {\doibase
  10.1103/PhysRevX.10.011047} {\bibfield  {journal} {\bibinfo  {journal} {Phys.
  Rev. X}\ }\textbf {\bibinfo {volume} {10}},\ \bibinfo {pages} {011047}
  (\bibinfo {year} {2020})}\BibitemShut {NoStop}%
\bibitem [{\citenamefont {Moudgalya}\ \emph {et~al.}(2022)\citenamefont
  {Moudgalya}, \citenamefont {Bernevig},\ and\ \citenamefont
  {Regnault}}]{Frag2}%
  \BibitemOpen
  \bibfield  {author} {\bibinfo {author} {\bibfnamefont {S.}~\bibnamefont
  {Moudgalya}}, \bibinfo {author} {\bibfnamefont {B.~A.}\ \bibnamefont
  {Bernevig}}, \ and\ \bibinfo {author} {\bibfnamefont {N.}~\bibnamefont
  {Regnault}},\ }\href {\doibase 10.1088/1361-6633/ac73a0} {\bibfield
  {journal} {\bibinfo  {journal} {Reports on Progress in Physics}\ }\textbf
  {\bibinfo {volume} {85}},\ \bibinfo {pages} {086501} (\bibinfo {year}
  {2022})}\BibitemShut {NoStop}%
\bibitem [{\citenamefont {Blackmore}\ \emph {et~al.}(2020)\citenamefont
  {Blackmore}, \citenamefont {Gregory}, \citenamefont {Bromley},\ and\
  \citenamefont {Cornish}}]{Blackmore}%
  \BibitemOpen
  \bibfield  {author} {\bibinfo {author} {\bibfnamefont {J.~A.}\ \bibnamefont
  {Blackmore}}, \bibinfo {author} {\bibfnamefont {P.~D.}\ \bibnamefont
  {Gregory}}, \bibinfo {author} {\bibfnamefont {S.~L.}\ \bibnamefont
  {Bromley}}, \ and\ \bibinfo {author} {\bibfnamefont {S.~L.}\ \bibnamefont
  {Cornish}},\ }\href {\doibase 10.1039/D0CP04651E} {\bibfield  {journal}
  {\bibinfo  {journal} {Phys. Chem. Chem. Phys.}\ }\textbf {\bibinfo {volume}
  {22}},\ \bibinfo {pages} {27529} (\bibinfo {year} {2020})}\BibitemShut
  {NoStop}%
\bibitem [{\citenamefont {Signoles}\ \emph {et~al.}(2014)\citenamefont
  {Signoles}, \citenamefont {Facon}, \citenamefont {Grosso}, \citenamefont
  {Dotsenko}, \citenamefont {Haroche}, \citenamefont {Raimond}, \citenamefont
  {Brune},\ and\ \citenamefont {Gleyzes}}]{Signoles2014}%
  \BibitemOpen
  \bibfield  {author} {\bibinfo {author} {\bibfnamefont {A.}~\bibnamefont
  {Signoles}}, \bibinfo {author} {\bibfnamefont {A.}~\bibnamefont {Facon}},
  \bibinfo {author} {\bibfnamefont {D.}~\bibnamefont {Grosso}}, \bibinfo
  {author} {\bibfnamefont {I.}~\bibnamefont {Dotsenko}}, \bibinfo {author}
  {\bibfnamefont {S.}~\bibnamefont {Haroche}}, \bibinfo {author} {\bibfnamefont
  {J.-M.}\ \bibnamefont {Raimond}}, \bibinfo {author} {\bibfnamefont
  {M.}~\bibnamefont {Brune}}, \ and\ \bibinfo {author} {\bibfnamefont
  {S.}~\bibnamefont {Gleyzes}},\ }\href {\doibase 10.1038/nphys3076} {\bibfield
   {journal} {\bibinfo  {journal} {Nature Physics}\ }\textbf {\bibinfo {volume}
  {10}},\ \bibinfo {pages} {715} (\bibinfo {year} {2014})}\BibitemShut
  {NoStop}%
\bibitem [{\citenamefont {Flo\ss{}}\ \emph {et~al.}(2015)\citenamefont
  {Flo\ss{}}, \citenamefont {Kamalov}, \citenamefont {Averbukh},\ and\
  \citenamefont {Bucksbaum}}]{floss2015observation}%
  \BibitemOpen
  \bibfield  {author} {\bibinfo {author} {\bibfnamefont {J.}~\bibnamefont
  {Flo\ss{}}}, \bibinfo {author} {\bibfnamefont {A.}~\bibnamefont {Kamalov}},
  \bibinfo {author} {\bibfnamefont {I.~S.}\ \bibnamefont {Averbukh}}, \ and\
  \bibinfo {author} {\bibfnamefont {P.~H.}\ \bibnamefont {Bucksbaum}},\ }\href
  {\doibase 10.1103/PhysRevLett.115.203002} {\bibfield  {journal} {\bibinfo
  {journal} {Phys. Rev. Lett.}\ }\textbf {\bibinfo {volume} {115}},\ \bibinfo
  {pages} {203002} (\bibinfo {year} {2015})}\BibitemShut {NoStop}%
\bibitem [{\citenamefont {Tuchendler}\ \emph {et~al.}(2008)\citenamefont
  {Tuchendler}, \citenamefont {Lance}, \citenamefont {Browaeys}, \citenamefont
  {Sortais},\ and\ \citenamefont {Grangier}}]{Tuchendler08}%
  \BibitemOpen
  \bibfield  {author} {\bibinfo {author} {\bibfnamefont {C.}~\bibnamefont
  {Tuchendler}}, \bibinfo {author} {\bibfnamefont {A.~M.}\ \bibnamefont
  {Lance}}, \bibinfo {author} {\bibfnamefont {A.}~\bibnamefont {Browaeys}},
  \bibinfo {author} {\bibfnamefont {Y.~R.~P.}\ \bibnamefont {Sortais}}, \ and\
  \bibinfo {author} {\bibfnamefont {P.}~\bibnamefont {Grangier}},\ }\href
  {\doibase 10.1103/PhysRevA.78.033425} {\bibfield  {journal} {\bibinfo
  {journal} {Phys. Rev. A}\ }\textbf {\bibinfo {volume} {78}},\ \bibinfo
  {pages} {033425} (\bibinfo {year} {2008})}\BibitemShut {NoStop}%
\bibitem [{\citenamefont {Walker}\ and\ \citenamefont
  {Saffman}(2012)}]{Walker2012}%
  \BibitemOpen
  \bibfield  {author} {\bibinfo {author} {\bibfnamefont {T.~G.}\ \bibnamefont
  {Walker}}\ and\ \bibinfo {author} {\bibfnamefont {M.}~\bibnamefont
  {Saffman}},\ }in\ \href@noop {} {\emph {\bibinfo {booktitle} {Advances in
  Atomic, Molecular, and Optical Physics}}},\ Vol.~\bibinfo {volume} {61}\
  (\bibinfo  {publisher} {Elsevier},\ \bibinfo {year} {2012})\ pp.\ \bibinfo
  {pages} {81--115}\BibitemShut {NoStop}%
\bibitem [{\citenamefont {Sibalic}\ \emph {et~al.}(2017)\citenamefont
  {Sibalic}, \citenamefont {Pritchard}, \citenamefont {Adams},\ and\
  \citenamefont {Weatherill}}]{sibalic17soft}%
  \BibitemOpen
  \bibfield  {author} {\bibinfo {author} {\bibfnamefont {N.}~\bibnamefont
  {Sibalic}}, \bibinfo {author} {\bibfnamefont {J.}~\bibnamefont {Pritchard}},
  \bibinfo {author} {\bibfnamefont {C.}~\bibnamefont {Adams}}, \ and\ \bibinfo
  {author} {\bibfnamefont {K.}~\bibnamefont {Weatherill}},\ }\href
  {https://www.sciencedirect.com/science/article/pii/S0010465517301972?via%3Dihub}
  {\bibfield  {journal} {\bibinfo  {journal} {Computer Physics Communications}\
  }\textbf {\bibinfo {volume} {220}},\ \bibinfo {pages} {319 } (\bibinfo {year}
  {2017})}\BibitemShut {NoStop}%
\end{thebibliography}%

\clearpage

\renewcommand{\thesection}{\Alph{section}}
\renewcommand{\thefigure}{S\arabic{figure}}
\renewcommand{\thetable}{S\Roman{table}}
\setcounter{figure}{0}
\renewcommand{\theequation}{S\arabic{equation}}
\renewcommand{\thepage}{S\arabic{page}}
\setcounter{equation}{0}
\setcounter{page}{1}

\newpage

\begin{widetext}
\appendix

\section{Supplemental Material for \\ ``Strongly interacting Rydberg atoms in synthetic dimensions with a magnetic flux''}

\vspace{5mm}

\section{Experimental initialization procedure}

\begin{figure*}[b]
	\includegraphics[width=0.99\textwidth]{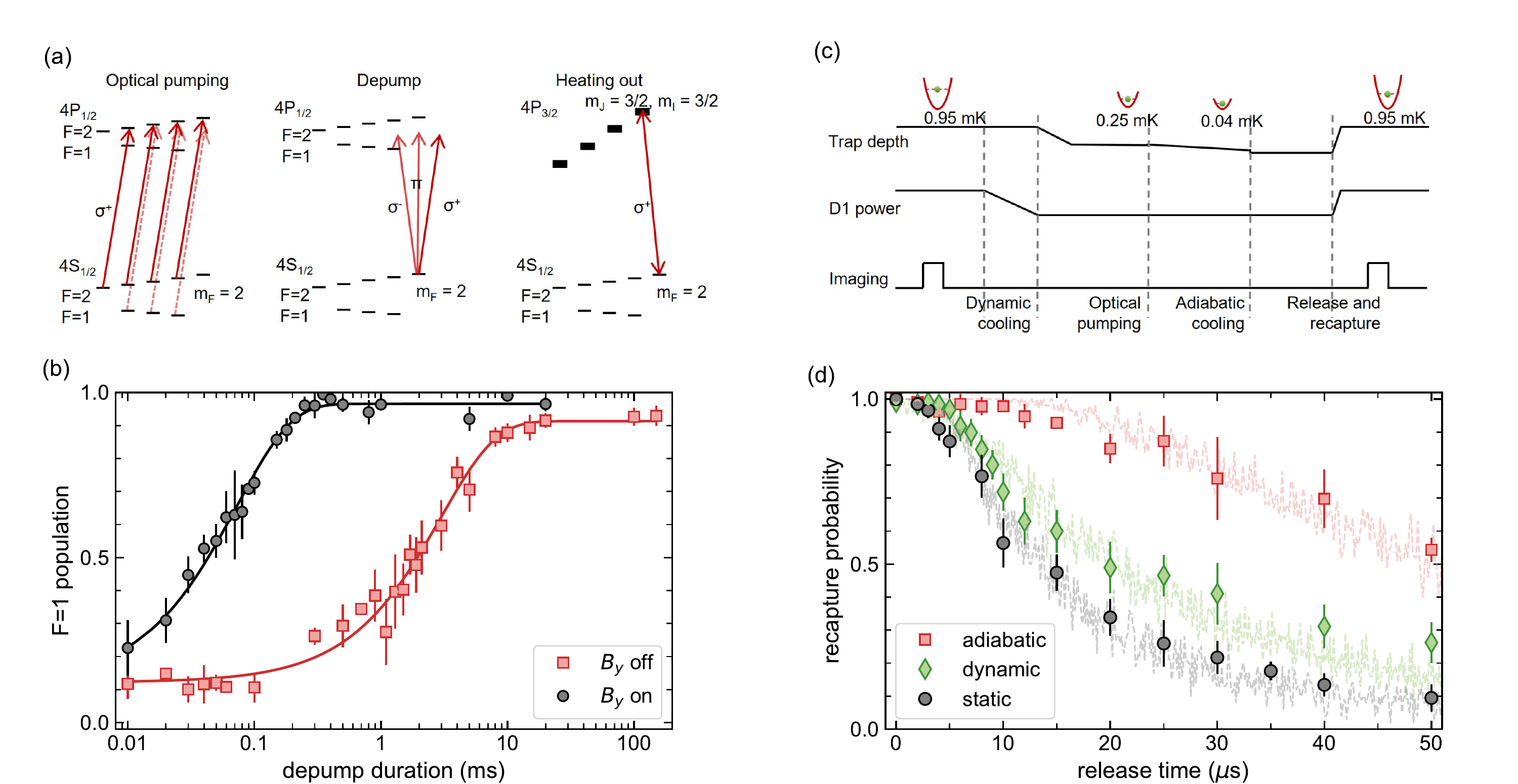}
    \caption{\textbf{Initial state preparation and temperature optimization procedure.}
    \textbf{(a)}~Depump-heating out sequence to optimize the $\sigma^+$ polarization of the D1 optical pumping beam. After pumping atoms into $|F=2, m_F=2\rangle$ dark state, we turn the repump component off, and the atoms get depumped into $|F=1\rangle$ states by the imperfect $\pi$ and $\sigma^-$ light in the optical pumping beam. Then we apply a heating beam (weak D2 light resonant to the $|F=2, m_F=2\rangle \leftrightarrow |4P_{3/2}, m_j = 3/2\rangle$ cycling transition) to remove $|F=2\rangle$ atoms. Measurement of the surviving atom number provides a measurement of the $|F=1\rangle$ population after the depumping step.
    %in $|F=1\rangle$ after a finite duration of depump process to resolve the depump time constant. 
    \textbf{(b)}~Measured $F=1$ population versus the depump duration for the case with only magnetic field along the $z$-direction (red square, $B_y$ off) and that with an additional magnetic field in the $y$-direction (black circle, $B_y$ on). Since $B_y$ changes the quantization axis, the atoms suffer from a relatively rapid depumping process. The exponential fittings (solid lines) give the corresponding time constant: $\tau_{\rm op}=72(2)~\mu {\rm s}$ with $B_y$ on and $\tau_{\rm dp}=3.0(2)~{\rm ms}$ with $B_y$ off. 
    \textbf{(c)}~Time sequence for optimization of the atom temperature with release-recapture measurement:  
(1) After first imaging with a trap depth of $\sim$0.95~${\rm mK}$, we dynamically ramp down the D1 power at a fixed trap depth, cooling the atoms from $\sim$60~$\mu{\rm K}$ (static cooling without changing D1 power) down to $\sim$20~$\mu{\rm K}$; 
(2) Then, we ramp the trap depth down to $\sim$0.25~${\rm mK}$ and perform optical pumping with D1 frequency shifted to compensate the revised trap light shift;
(3) We further adiabatically ramp the trap depth down to $\sim$40~$\mu{\rm K}$ to further cool the atoms to $\sim$4~$\mu{\rm K}$. 
    \textbf{(d)}~Release and recapture measurements of the atom temperature after different cooling processes. The temperatures are resolved by fitting the respective experimental data sets to Monte-Carlo simulations (dashed lines).
}
\label{FIG:figs1}
\end{figure*}

We begin our experiments by loading $^{39}$K atoms into one-dimensional optical tweezer arrays generated by diffraction of 780~nm laser light from an acousto-optic deflector (AA Opto-Electronic part number DTSX-400-780). Every cycle of the experiment (having a duration of 1.3~s), atoms are first probabilistically loaded into the tweezer traps with an average probability of $\sim$55$\%$. Two different tweezer patterns are used for the studies presented: the pattern of seven two-site dimers depicted in Fig.~\pref{FIG:fig1}{a,b} as well as a pattern of three six-site clusters used for the data in Fig.~\ref{FIG:fig4}.

After an initial loading, the samples of atoms are nondestructively imaged a first time for subsequent post-selection. The fluorescence imaging, with a duration of 40~ms, is characterized by a survival probability $>$99$\%$ and a discrimination fidelity (for the assignment of an atom occupancy or vacancy) $>$99$\%$~\cite{JacksonPRR}.
The atoms are then re-cooled by gray molasses~\cite{JacksonPRR,Gross-SciPost} as well as adiabatic trap decompression (lowering the depth of our Gaussian tweezer traps from an initial depth of $\sim$0.95~mK to a final depth of $40$~$\mu$K) to a final temperature of $\sim$4~$\mu$K as calibrated by release and recapture~\cite{Tuchendler08}. This overall preparation procedure, as well as the release-and-recapture probability curves (and associated numerical comparisons for temperature estimation) can be seen in Fig.~\pref{FIG:figs1}{c,d}. The three release-and-recapture curves in Fig.~\pref{FIG:figs1}{d} are respectively characterized by a temperature of $\sim$60~$\mu{\rm K}$ (black data points and theory) under static D1 cooling in the trap at its initial depth, a reduced temperature of $\sim$20~$\mu{\rm K}$ (green data points and theory) following a dynamical reduction of the D1 molasses cooling power, and a minimum temperature of $\sim$4~$\mu{\rm K}$ (red data points and theory) following an adiabatic ramp down of the optical tweezer trap to a final depth of $\sim$40~$\mu{\rm K}$.

In preparation for the Rydberg studies, prior to the final adiabatic cooling stage, the trapped atoms are optically pumped to a single ground internal state, $\ket{4 S_\textrm{1/2}, F = 2, m_F = 2}$, with $\sim$98(1)$\%$ efficiency. As depicted in Fig.~\pref{FIG:figs1}{a,b}, the efficiency is estimated by comparing the characteristic depumping time for the case of the actual bias field $B_z=27~{\rm G}$ implemented in experiment ($\tau_{\rm dp}=3.0(2)~{\rm ms}$, $B_y$ off) to the case where an additional magnetic field is added along the $y$ axis ($\tau_{\rm op}=72(2)~\mu {\rm s}$, $B_y$ on, with $B_y\sim 12~{\rm G}$).
This additional field $B_y$ disrupts the polarization purity of the depumping beam relative to the total quantization axis, and the corresponding measurement provides a lower estimate for the depumping rate for unpolarized light. The depumping times for these two situations can be combined to estimate the optical pumping (OP) efficiency $\eta = 1- \tau_{\rm op}/\tau_{\rm dp} = 98(1)\%$~\cite{Walker2012}.

\begin{figure*}[b]
	\includegraphics[width=0.95\textwidth]{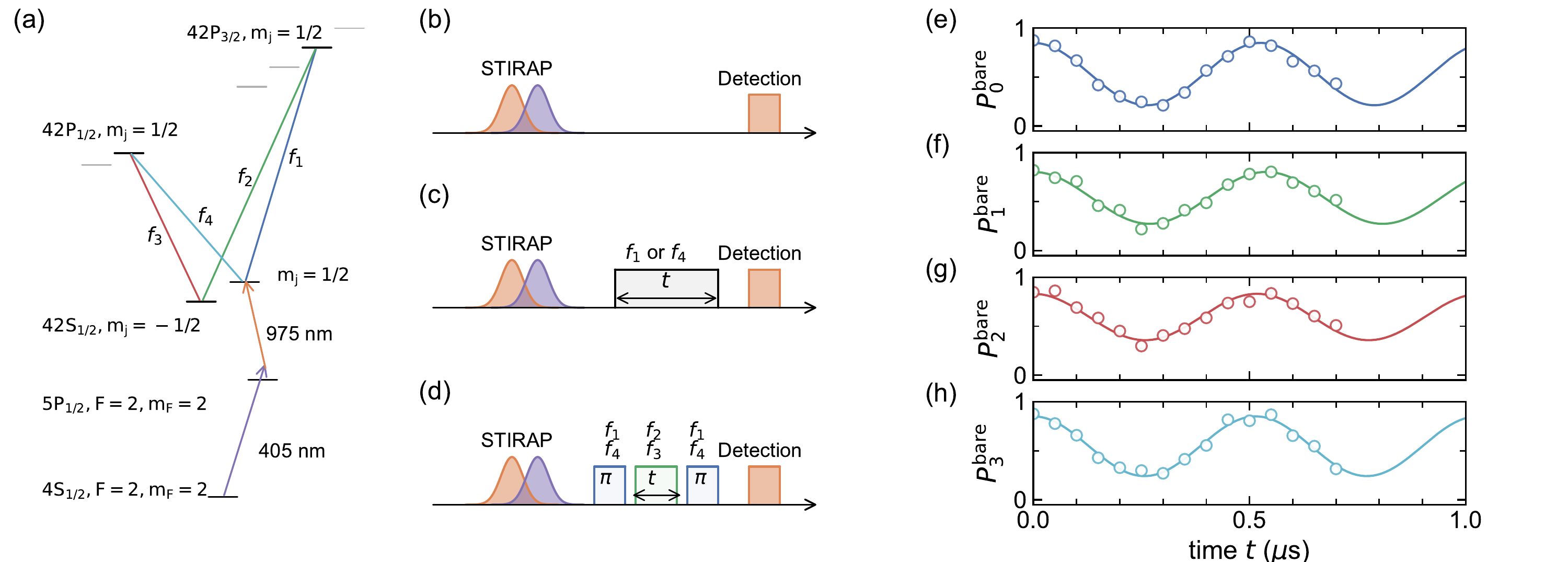}
	\caption{\textbf{Calibration of the Rabi frequencies between each Rydberg state pair in isolated single atoms.}
        \textbf{(a)}~Level structure used in our experiment.
        \textbf{(b)}~Time sequence with only STIRAP pulses and 975 nm de-excitation pulse to calibrate the baselines of the oscillation between Rydberg state pairs.
        \textbf{(c)}~Time sequence with an extra MW pulse to measure the Rabi oscillation between $|0\rangle = |42S_{1/2}, m_j=1/2\rangle$ and $|1\rangle=|42P_{3/2}, m_j=1/2\rangle$ ($|3\rangle = |42P_{1/2}, m_j=1/2\rangle$) driven by $f_1$ ($f_4$). 
        \textbf{(d)}~Time sequence with two extra $\pi$-pulses to measure the Rabi oscillation between $|2\rangle = |42S_{1/2}, m_j=-1/2\rangle$ and $|1\rangle=|42P_{3/2}, m_j=1/2\rangle$ ($|3\rangle = |42P_{1/2}, m_j=1/2\rangle$) driven by $f_2$ ($f_3$). 
        \textbf{(e-h)}~Measured uncorrected state population dynamics when driving pairwise Rabi oscillations between states addressed by the tones $f_1$ to $f_4$, respectively (from top to bottom). The solid line fits are damped sine functions used to calibrate the respective Rabi rates we use as $\{\Omega_{01}, \Omega_{12}, \Omega_{23}, \Omega_{30}\}/h = \{1.90(3), 1.86(4), 1.93(4), 1.94(4)\}~{\rm MHz}$. 
}
\label{FIG:figs2}
\end{figure*}

The excitation of the atoms to Rydberg levels (principal quantum number $n = 42$) is performed after releasing the atoms from the optical tweezer traps, which are weakly anti-trapping (with a polarizability that is roughly 30 times lower in magnitude as compared to that for the ground state) for the target Rydberg level. After $\sim$0.2~$\mu$s of release, a two-photon STIRAP pulse is applied, as depicted in Fig.~\ref{FIG:figs2}. This excitation involves laser light slightly ($\sim$15~MHz) detuned from the $\ket{4 S_\textrm{1/2}, F = 2, m_F = 2} \leftrightarrow \ket{5 P_\textrm{1/2}, F = 2, m_F = 1/2}$ transition (``lower leg,'' having a wavelength of $\sim$405~nm) and the $\ket{5 P_\textrm{1/2}, F = 2, m_F = 1/2} \leftrightarrow \ket{42 S_\textrm{1/2}, m_J = 1/2}$ transition (``upper leg,'' having a wavelength of $\sim$975~nm). The two lasers used for STIRAP are stabilized via Pound-Drever-Hall (PDH) locking to a common ultra-low expansion (ULE) optical cavity (Stable Laser Systems). Peak single-photon (resonant) Rabi rates of $\sim$$2\pi \times 20$~MHz for the lower and upper legs of the STIRAP transition are achieved by focusing the combined laser beams to respective waists of 40~$\mu$m and 30~$\mu$m. The peak powers $I_{0, L/U}$ at the atoms are roughly 3~mW for the ``lower leg'' and 300~mW (after amplification by a tapered amplifier) for the ``upper leg,'' respectively.
We use Gaussian-shaped pulses for both the lower and upper legs, as shown in Fig.~\pref{FIG:figs2}{b}, the intensities of which follow the formula $I_{L/U}(t) = I_{0, L/U} {\rm exp}\left[-\frac{(t\pm \Delta t/2)^2}{\sigma^2}\right]$ with $\Delta t=0.3~\mu$s and $\sigma = 0.3~\mu$s.
Under these conditions, we achieve a one-way STIRAP efficiency of $\sim$94(1)~$\%$.

\section{The synthetic lattice -- state configuration and calibration}

The mapping between bare Rydberg levels and the sites of our synthetic lattice are detailed in the main text Fig.~\pref{FIG:fig1}{c,e}. These chosen assignments are informed by a few simple considerations: (i)~we would like all of the state-to-state transitions to be achieved by dipole-allowed first-order processes, (ii)~we desire for the resonant exchange interactions to only occur between nearest neighbors in the synthetic dimension, and (iii)~for technical considerations, we require that all of the transitions can be addressed with only a moderate microwave bandwidth. The ``folded diamond'' layout of Fig.~\pref{FIG:figs2}{a} satisfies these design goals.

\begin{figure*}[b]
	\includegraphics[width=0.91\textwidth]{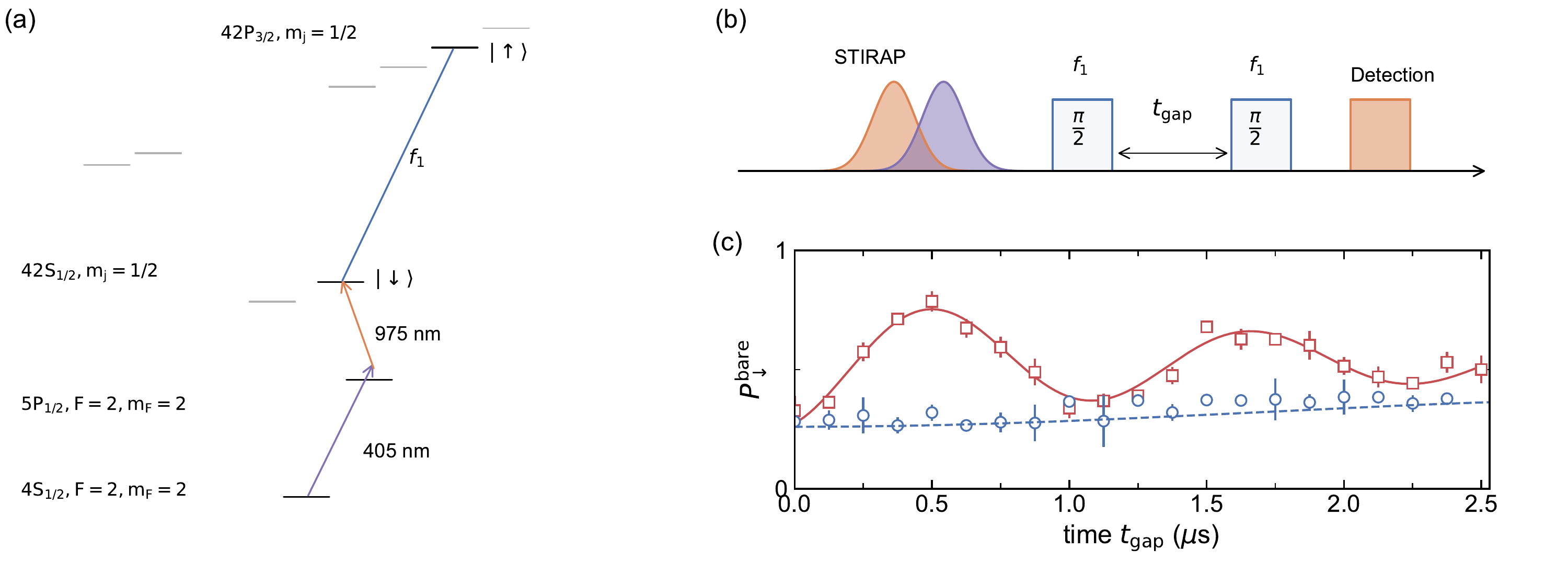}
	\caption{\textbf{Calibration of the dipole-dipole interaction strength.}
        \textbf{(a)}~Level structure used to measure the dipolar exchange interaction with $\ket{\downarrow} = |0\rangle = |42S_{1/2}, m_j=1/2\rangle$ and $\ket{\uparrow} = |1\rangle = |42P_{3/2}, m_j=1/2\rangle$. The interatomic distance is set to $\approx 10~\mu{\rm m}$, at exactly twice the spacing as used for the experiments described in the main text.
        \textbf{(b)}~Timing diagram of the sequence used to measure Ramsey coherence oscillations for pairs of interacting atoms. After initially preparing a pair of atoms in $\ket{\downarrow\downarrow}$ via STIRAP, a strong microwave $\pi/2$ pulse (Rabi frequency $\sim$$2\pi\times 4.0$~MHz) is applied to rotate the atom pair to a product state $\ket{\downarrow + i \uparrow}\ket{\downarrow + i \uparrow}/2 = [\ket{\downarrow\downarrow}-\ket{\uparrow\uparrow}+ i (\ket{\uparrow\downarrow}+\ket{\downarrow\uparrow})]/2$. Then we let the system freely evolve for a duration of $t_{\rm gap}$, during which the time evolution follows $[\ket{\downarrow\downarrow}-\ket{\uparrow\uparrow}+ i e^{iV_{01}^{\rm cal}t_{\rm gap}/\hbar}(\ket{\uparrow\downarrow}+\ket{\downarrow\uparrow})]/2$ as the state $\ket{\uparrow\downarrow}+\ket{\downarrow\uparrow}$ has an eigenenergy of $V_{01}^{\rm cal}$. We finally read out the population in $\ket{\downarrow}$ after applying an identical microwave $\pi/2$ pulse and 975~nm de-excitation pulse.
        \textbf{(c)}~Detected population in $\ket{\downarrow}$ state versus the free evolution time $t_{\rm gap}$ for non-interacting single atoms (blue circles) and for the atoms of interacting pairs (red squares). The fitting (solid line) to data for atom pairs with a damped sine function gives the oscillation frequency, \textit{i.e.}, the interaction strength, $V_{01}^{\rm cal}/h = 0.86(2)~{\rm MHz}$. In contrast, fitting (dashed line) to the single-atom data with a sine function shows only a slow variation, indicating the frequency of microwave $f_1$ is detuned from resonance by $\sim$150~${\rm kHz}$.
}
\label{FIG:figs3}
\end{figure*}

The microwave transition frequencies between each Rydberg state pair in our system lie in the vicinity of 48~GHz, as shown in Fig.~\pref{FIG:figs2}{a}. We first input a single-tone microwave signal at $\sim$12~GHz (generated from a Vaunix Lab Brick device) into a $4\times$ frequency multiplier (Marki AQA-2156). We then mix this high-frequency carrier signal with two multi-tone arbitrary waveform signals (generated from a Teledyne SDR14TX card) via an IQ frequency mixer (Marki MMIQ-4067L) to produce the required sidebands to resonantly couple the relevant pairs of states in our system.

Readout from the various Rydberg levels is achieved through combinations of microwave state-swapping pulses and optical depumping on the ``upper leg'' 975~nm transition. For the $\ket{0}$ state that we initially populate via STIRAP, the readout for measuring the population $P_0$ simply involves applying near-resonant depumping on the upper leg  975~nm transition (followed by ground state imaging). To note, depumping from the nearby $\ket{2}$ state (75~MHz away in energy at these moderate bias fields) is avoided by using sufficiently weak intensities of the 975~nm depumping light. As described in the main text, the ability to depump both the $\ket{0}$ and $\ket{2}$ state simultaneously by applying high-intensity depump light provides a useful way to measure the $\ket{2}$ state population as $P_2 = P_{0+2} - P_0$. For the states $\ket{1}$ and $\ket{3}$, relating to $42P$ levels, the populations $P_1$ and $P_3$ can be read out by applying $\pi$ pulses on the $\ket{0} \leftrightarrow \ket{1}$ or the $\ket{0} \leftrightarrow \ket{3}$ transition prior to measurement of the $\ket{0}$ state population. And, while not utilized in this study, coherences between the sites in the synthetic dimension can also be read out in such a way.

Figure~\ref{FIG:figs2} details the procedure for calibrating the effective tunneling rates (transition Rabi rates) along the various links of the lattice. Because the various transitions involve different sets of states and require different polarizations of microwaves, the individual amplitudes of the tones $f_{1-4}$ are adjusted to achieve a uniform tunneling rate across all links.
%Calibration of the Rabi rates, as depicted in Fig.~\pref{FIG:figs1}{e-h}.
As described in the text, the overall flux $\phi$ of the synthetic diamond lattice is calibrated based on the dynamical response of isolated single atoms (cf. Fig.~\ref{FIG:fig3}) and its comparison to theory. The value of $\phi$ is set by controlling the source phase of the $f_1$ frequency tone relative to the other tones.
%Calibrated based on the comparison of the response of singles to the predictions of theory for a simple diamond lattice (see Fig.~\ref{FIG:fig3}).
To note, we find in experiment that the flux $\phi$ is extremely stable, with no noticeable variations on the week-long timescales hitherto explored.

\begin{figure*}[b]
	\includegraphics[width=0.95\textwidth]{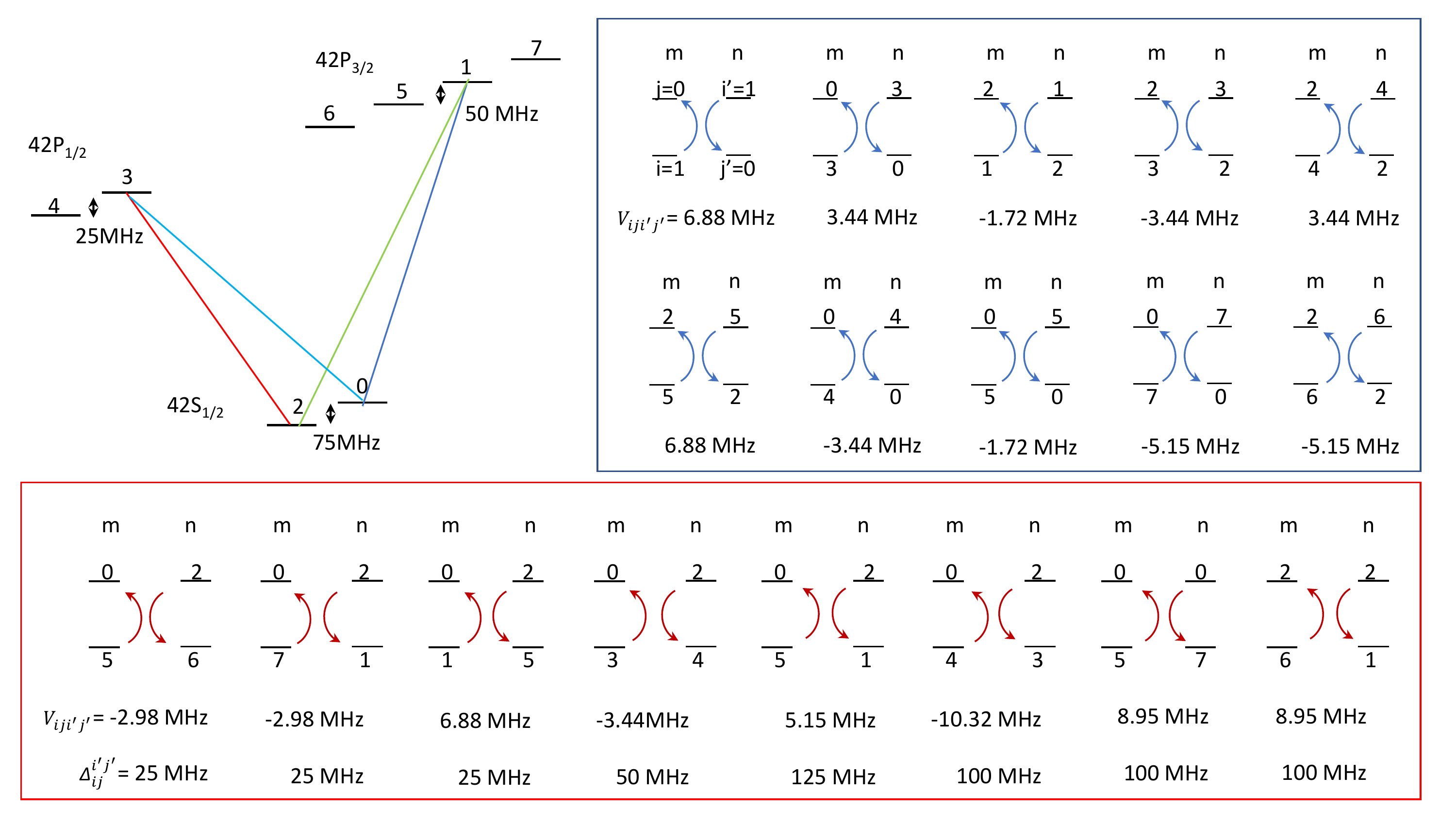}
    \caption{\textbf{All dipole-dipole interaction terms relevant to our experimental scheme.}
    Blue box: Resonant flip-flop interactions with $\Delta_{ij}^{i'j'} = 0$. We list the interaction energies (reduced by Planck's constant) in MHz.
    Red box: Non-resonant state-changing interaction terms. The listed interaction strength $V_{iji'j'}$ values (also reduced by Planck's constant and listed in MHz) are based on the experimentally calibrated $V_{01}=6.88~{\rm MHz}$ scaled by the respective $C_3$ coefficients calculated with the Alkali Rydberg calculator (ARC) package~\cite{sibalic17soft}. We also list the configurational detuning $\Delta_{ij}^{i'j'}$ with respect to our synthetic lattice states for each of the processes (also in MHz).
}
\label{FIG:figs4}
\end{figure*}

\section{Dipolar interactions -- theoretical expectation and experimental calibration}

In the language of our tight-binding ``synthetic lattice'' model, the dipole-dipole interaction Hamiltonian is

\begin{equation}
 H_{\rm int} = \sum_{m,n} \sum_{i,j,i',j'}
 V_{iji'j'}^{mn}e^{i\Delta_{ij}^{i'j'}t}
 \hat{c}_{i,m}^\dagger \hat{c}_{j,m} \hat{c}_{j',n}^\dagger \hat{c}_{i',n} + {\rm h.c.} \ ,
\end{equation}
where $V_{iji'j'}^{mn} = \langle j_m i'_n | V_{\rm dd} | i_m j'_n\rangle$ with the dipolar interaction operator $V_{\rm dd} = \frac{1}{4\pi\epsilon_0 R^3_{mn}}\left[\frac{1}{2}(2d_m^0d_n^0 + d_m^+d_n^- + d_m^-d_n^+) - \frac{3}{2}(d_m^-d_n^- + d_m^+d_n^+)\right]$ between atom $m$ and atom $n$ ($d^0, d^+$ and $d^-$ are the respective dipole moment operators for $\pi$, $\sigma^+$ and $\sigma^-$ transitions), and where $\Delta_{ij}^{i'j'}$ is the energy difference between $|i_m \rangle \leftrightarrow |j_m\rangle$ and $|i'_n\rangle \leftrightarrow |j'_n\rangle$ transitions (or, equivalently, the energy difference between the two-body state configurations $|i_m\rangle |j'_n\rangle$ and $|j_m\rangle |i'_n\rangle$). Here the state index $i,j~(i',j')$ also covers other unused sublevels in both $42P$ manifolds, as displayed in Fig.~\ref{FIG:figs4}, to include the strong state-changing dipolar interactions.

There are four primary dipolar exchange processes that we care about (\textit{i.e.}, they are the only four processes that are resonant for the states intentionally populated in our experiment), occurring between pairs of atoms occupying the states $\ket{0}$ and $\ket{1}$ (with an energy scale $V_{01}$), $\ket{1}$ and $\ket{2}$ ($V_{12}$), $\ket{2}$ and $\ket{3}$ ($V_{23}$), and $\ket{3}$ and $\ket{0}$ ($V_{30}$). Based on the construction of our synthetic lattice, all of these resonant ``flip-flop'' terms occur between pairs of atoms residing on neighboring sites of our synthetic lattice. Importantly, in this work the population dynamics is also impacted by the presence of relatively strong state-changing dipolar interactions that are not very far off from resonance (because we operate with only a moderate quantization field). The full enumeration of resonant population-conserving ($\Delta \ell = 0$) and off-resonant population non-conserving ($\Delta \ell = \pm 2$) dipolar interaction terms are presented in Fig.~\ref{FIG:figs4}. The resonant terms are listed in the blue box, and all relevant (not off-resonant by more than 125~MHz) non-resonant exchange terms are enumerated in the red box. As described in the main text, simulation comparisons generally include both the idealized interaction scenarios (resonant only, dashed lines) as well as the full expectations based on textbook dipolar physics (all terms, solid lines).

Based on our imaged tweezer patterns and the designed magnification of our imaging system, we expect our tweezer trap spacing to be $\sim$5~$\mu$m. Based on this spacing and the known~\cite{sibalic17soft} $C_3$ coefficients for the Rydberg states we consider, we can obtain estimates for these various dipolar interaction energies. Because the imaging system's magnification is not independently calibrated, however, we perform direct measurements of the dipolar exchange rates as a primary calibration of the dipolar interaction energies.
First, we have measured the energy detuning of the triplet resonance from the single-atom resonance for the terms $V_{01}$ and $V_{30}$. These measurements are performed in the actual ($\sim$5~$\mu$m spacing) tweezer configuration utilized in the main text. The results confirm the relative magnitude and signs of the expected $V_{ij}$ exchange energies, and in combination these measurements suggest a tweezer spacing of $5.1(2)$~$\mu$m.
However, we seek an alternative and more sensitive calibration of the dipolar exchange rate, using the technique of Ramsey coherence oscillations as employed in Ref.~\cite{Yan2013}.
The procedure for this is depicted in Fig.~\ref{FIG:figs2}. This measurement involves performing high-fidelity $\pi/2$ rotations before and after some free evolution period ($t_\textrm{gap}$). Because the achievement of high-fidelity $\pi/2$ pulses is difficult in the presence of strong interactions, and hence for our 5~$\mu$m tweezer spacing, we in fact perform this calibration measurement in an array that has exactly twice the spacing between neighboring atom traps (as set by the frequency tones applied to the acousto-optic deflector). We first tune the transition frequency very close to resonance, such that for singles (empty blue circles) we only observe a slow variation of the Ramsey signal (due to a slight precession of the spin during the Ramsey gap time, as the second $\pi/2$ pulse has no phase shift relative to the first pulse). For pairs, we observe an additional oscillation of the average Ramsey coherence, consistent with the coherent entangling and disentangling of the atoms at the rate $V_{01}/h$. This measured rate of $V^\text{cal}_{01}/h = 0.86(2)$~MHz is 4 times smaller than the value of $V$ as defined previously, setting $V/h = 3.44(8)$~MHz for our short-spacing arrays and likewise confirming a spacing of 4.8(1)~$\mu$m based on comparison to the predictions of the Alkali Rydberg calculator (ARC) package~\cite{sibalic17soft}.

\section{Corrections for preparation and readout infidelity}

The primary data we measure for all state populations $P_{0-3}$ appear similar to those presented in Fig.~\ref{FIG:figs2} and Fig.~\ref{FIG:figs3}. There are two limiting quantities to note. First, there is an upper baseline value that is on average equal to $P_u = 0.88(1)$, which stems from inefficiencies of STIRAP and release-and-recapture survival. There is also a lower baseline of the measurements, having a value $P_l = 0.21(1)$, that we believe stems from the decay (and subsequent recapture) of the short-lived $n = 42$ Rydberg states. This lower baseline represents a lack of fidelity in discriminating atoms from being successfully depumped from the state of choice as opposed to decaying from any of the Rydberg levels. These infidelities limit the contrast of single atom dynamics, and more importantly limit our ability to faithfully measure atom-atom correlation dynamics.
%Because we operate with relatively low-lying Rydberg levels ($n = 42$), the decay of the Rydberg states to the ground state during the free release and recapture period leads to a background ``recapture'' signal of $20\%$ (\textit{i.e.}, even for states that were not resonantly depumped).

For all of the data in the paper, we ``correct'' for these known infidelities in the following way: we define the corrected populations $P_i$ in relation to the measured bare populations $P_i^\textrm{bare}$ as $P_i = (P_i^\textrm{bare}-P_l)/(P_u-P_l)$.
%
%
%The data for the state populations has been ``corrected'' for this fixed background signal, as well as the $95\%$ STIRAP efficiency, as $P_i = (P_i^\textrm{bare}-P_l)/(P_u-P_l)$. Here, for the fractional population at the synthetic lattice site $\ket{i}$, $P_i$ and $P_i^\textrm{bare}$ are the plotted and bare measurement fractions. 

\begin{figure*}[b]
	\includegraphics[width=0.95\textwidth]{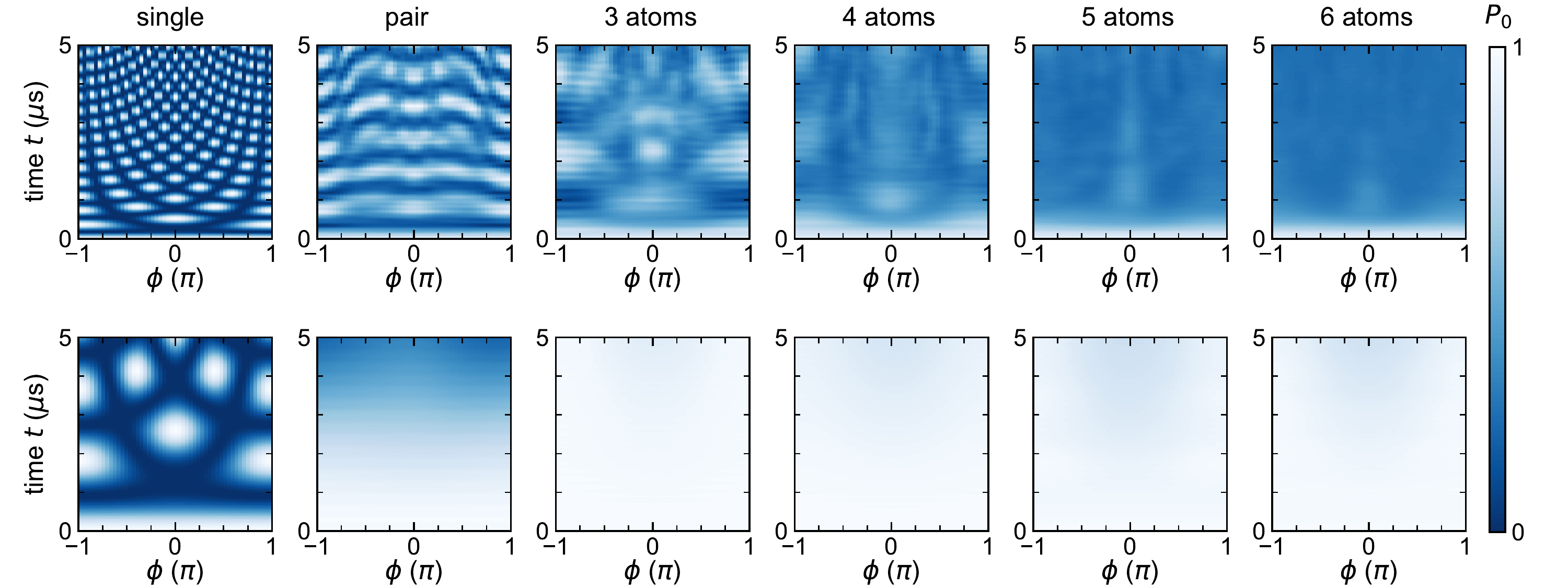}
    \caption{\textbf{$P_0$ dynamics for increasing lengths of few-atom clusters (left to right), for both intermediate (top, $V/\Omega = 1.8$) and strong (bottom, $V/\Omega = 9$) interactions.}
    Each plot shows the phase dependence of the array-averaged $P_0$ dynamics.
Top: $V/\Omega = 1.8$ and $\Omega/h = 1.92$~MHz, for atom array lengths from one to six.
Bottom: $V/\Omega = 9$ and $\Omega/h = 0.38$~MHz, for atom array lengths from one to six. Here, only the resonant flip-flop (state conserving) dipolar interactions are considered.
%{JPC: Consider showing the evolution for the same duration in units of $\Omega$ between the top and bottom. The top shows a much ``longer" effective timescale.}
}
\label{FIG:figs5}
\end{figure*}

\section{Interactions in few-atom arrays -- scrambled and frozen dynamics}

In Fig.~\ref{FIG:figs5}, we provide slightly more numerical evidence for the suggestive claims made in the main text that in our six-atom clusters we begin to see the emergence of ergodic dynamics and frozen dynamics in the regimes of intermediate and strong interactions. In the upper row of plots, we investigate over timescales of 0 to 5~$\mu$s the flux-dependent dynamics expected for atom arrays of varying size for the case of intermediate interactions ($V = 1.8 \Omega$, $\Omega/h = 1.92$~MHz). In this regime, with several interaction energies $V_{ij}$ being nearly on the same scale as the single-particle hopping terms, one may reasonably expect that the nonequilibrium dynamics of few-atom clusters becomes quite complex, with the absence of any revivals or oscillatory dynamics on reasonable timescales. This is what is observed in the six-atom calculations, where at reasonably short timescales of just a few $\mu$s there is essentially no flux dependence or dynamics to the $P_0$ measure, with a static value of $\approx 1/4$ found for all $\phi$ values. These numerical results suggest that nearly ergodic behavior may be expected in these interacting many-state systems.

In the lower row of plots of Fig.~\ref{FIG:figs5}, we instead show the flux-dependent dynamics (over the same timescale of 5~$\mu$s) that is expected for atom arrays in the strong interaction regime ($V = 9.0 \Omega$, $\Omega/h = 0.38$~MHz). In this regime, one sees that the addition of more and more atoms to the array has a very different influence on the dynamics. For pairs, the dynamics slows considerably as compared to singles and the case of pairs with intermediate interactions. For arrays with three or more atoms, the dynamics appears to nearly cease over the timescale investigated. As discussed in the main text, this is consistent with the expectation that zero-energy strings should become immobile if they are far separated in energy from the interacting configurations that would be populated by uncorrelated atom hopping in the synthetic dimension.
%The primary data we measure for all state populations $P_{0-3}$ appear similar to those presented in Fig.~\ref{FIG:figs2} and
\end{widetext}
\end{document}